\documentclass{article}
\usepackage{epsfig}
\usepackage{cite}
\def\bom#1{{\mbox{\boldmath $#1$}}}
\def\MSbar{$\overline{{\rm MS}}$}
\def\lapprox{\lower .7ex\hbox{$\;\stackrel{\textstyle <}{\sim}\;$}}
\def\gapprox{\lower .7ex\hbox{$\;\stackrel{\textstyle >}{\sim}\;$}}

\def\e{\epsilon}

\def\S{\, {\rm S}}
\def\G{{\rm G}}
\def\H{{\rm H}}

\def\CA{C_A}

\def\NF{N_F}
\def\NFZ{N_{F,V}}
\def\dd{d}
\def\dmb{(d-2)}
\def\dmc{(d-3)}
\def\dmd{(d-4)}
\def\dmf{(d-6)}
\def\dmj{(d-10)}
\def\sab{s_{12}}
\def\sac{s_{13}}
\def\sbc{s_{23}}

\def\sabc{s_{123}}
\def\Taa{T^\dagger_{11}\cdot \epsilon^*_4}
\def\Tab{T^\dagger_{12}\cdot \epsilon^*_4}
\def\Tac{T^\dagger_{13}\cdot \epsilon^*_4}
\def\Tba{T^\dagger_{21}\cdot \epsilon^*_4}
\def\Tbb{T^\dagger_{22}\cdot \epsilon^*_4}
\def\Tbc{T^\dagger_{23}\cdot \epsilon^*_4}
\def\T{T^\dagger\cdot \epsilon^*_4}
\def\S{{\cal S}}

\def\Bab{{\rm Bub}(\sab)}
\def\Bac{{\rm Bub}(\sac)}
\def\Bbc{{\rm Bub}(\sbc)}
\def\Babc{{\rm Bub}(\sabc)}
\def\bab{\left[\Bab-\Babc\right]}
\def\bac{\left[\Bac-\Babc\right]}
\def\bbc{\left[\Bbc-\Babc\right]}

\def\Boxxy{{\rm Box}^6(\sab,\sac,\sabc)}
\def\Boxyz{{\rm Box}^6(\sac,\sbc,\sabc)}
\def\Boxxz{{\rm Box}^6(\sab,\sbc,\sabc)}

\begin{document}
\unitlength1cm
\begin{titlepage}
\vspace*{-1cm}
\begin{flushright}
IPPP/02/28\\
DCTP/02/56\\
CERN-TH/2002-122\\
hep-ph/0206067\\
June 2002 
\end{flushright}                                
\vskip 2.5cm

\begin{center}
\boldmath
{\Large\bf Two-Loop QCD Helicity Amplitudes for  $e^+e^- \to 3$~Jets}
\unboldmath
\vskip 1.cm
{\large L.W.~Garland}$^a$, {\large T.~Gehrmann}$^b$, 
{\large E.W.N.~Glover}$^a$, {\large A.~Koukoutsakis}$^a$
and {\large E.~Remiddi}$^c$ 
\vskip .7cm
{\it $^a$ Department of Physics, University of Durham, Durham DH1 3LE, England}
\vskip .4cm
{\it $^b$ Theory Division, CERN, CH-1211 Geneva 23, Switzerland}
\vskip .4cm
{\it $^c$ Dipartimento di Fisica,
    Universit\`{a} di Bologna and INFN, Sezione di 
    Bologna,  I-40126 Bologna, Italy} 
\end{center}
\vskip 2.0cm

\begin{abstract}
We compute the two-loop QCD helicity amplitudes for the process 
$e^+e^- \to q\bar q g$. The amplitudes are extracted in a scheme-independent
manner from the coefficients appearing in the 
general tensorial structure for this process. The tensor coefficients are 
derived from the Feynman graph amplitudes
by means of projectors, within the
conventional dimensional regularization scheme.
The actual calculation of the 
loop integrals is then performed 
by reducing all of them
to a small set of known master integrals. 
The infrared pole structure of the renormalized helicity amplitudes 
agrees with the
prediction made by Catani using an infrared factorization formula. 
We use this formula to structure our results for 
the finite part into terms arising from 
the expansion of the pole coefficients and a genuine finite remainder,
which is independent of the scheme used to define the helicity amplitudes. 
The 
analytic result for the finite parts of 
the amplitudes is expressed 
in terms of one- and two-dimensional harmonic polylogarithms.

\end{abstract}
\vskip 3cm
\begin{tabbing}
{\it Email-addresses: }\= {\tt L.W.Garland@durham.ac.uk},
{\tt Thomas.Gehrmann@cern.ch},\\ \> {\tt E.W.N.Glover@durham.ac.uk},
{\tt Athanasios.Koukoutsakis@durham.ac.uk},\\ 
\>{\tt Ettore.Remiddi@bo.infn.it}
\end{tabbing}
\vfill

\end{titlepage}                                                                
\newpage

\renewcommand{\theequation}{\mbox{\arabic{section}.\arabic{equation}}}

\section{Introduction}
\setcounter{equation}{0}
The three-jet production rate in electron--positron 
annihilation~\cite{petra,ellis} 
and related event shape observables were measured to a very high 
precision at LEP, where they were used in particular for the 
determination of the strong coupling constant $\alpha_s$. At present,
the error on the extraction of
$\alpha_s$ from these  data is dominated by the uncertainty inherent in the
theoretical  next-to-leading order (NLO)  calculation~\cite{ert1,ert2,kn,gg,cs}
of the jet observables (see~\cite{bethke} for a review). 
The planned  TESLA~\cite{tesla} linear $e^+e^-$
collider will allow precision  QCD studies at energies even higher than at
LEP. Given the projected  luminosity of TESLA, one again expects the
experimental errors to  be well below the uncertainty of the NLO calculation. 

The calculation of  next-to-next-to-leading order (NNLO), i.e.\  ${\cal
O}(\alpha_s^3)$, corrections to the three-jet rate in $e^+e^-$ annihilation 
has been considered as a highly important project for a long 
time~\cite{kunszt}. In terms of matrix elements, it requires the computation
of three contributions: the tree level $\gamma^* \to 5$~partons
amplitudes~\cite{tree5p1,tree5p2,tree5p3}, the one-loop corrections to the $\gamma^* \to
4$~partons amplitudes~\cite{onel4p1,onel4p2,onel4p3,onel4p4},  and the two-loop
(as well as the one-loop  times one-loop) corrections to the  $\gamma^* \to
3$~partons matrix  elements. In a previous publication~\cite{3jme}, we have 
derived both the
interference of the tree and two-loop matrix elements and the self-interference
of the one-loop amplitudes averaged over all external helicities. 
In the present work, we extend this calculation to 
compute the two-loop helicity amplitudes for the process 
 $e^+e^- \to q\bar q g$.

The most precisely measured observables related to $e^+e^- \to 3$~jets
are the jet production rate itself
and a number of event-shape variables.
The calculation of these phenomenologically most relevant applications,
which also dominate the extraction of $\alpha_s$,
at NNLO accuracy 
requires only the helicity averaged squared matrix element at the 
two-loop level derived in~\cite{3jme}. Nevertheless, the helicity 
amplitudes presented here are interesting for a number of reasons:
\begin{itemize}
\item Oriented event-shape observables, which measure the spatial 
orientation of the final-state jets relative to the direction of 
the incoming beams require, even for 
unpolarized beams~\cite{osland}, 
the calculation of the polarization tensor of the 
virtual photon mediating the interaction. This polarization tensor 
can be recovered from the helicity amplitudes. 
\item Likewise, to determine the direction of the decay leptons in 
the crossed process, $V+1$ jet production at unpolarized hadron 
colliders, it is necessary to compute the polarization tensor of the 
vector boson. 
\item Polarization of the beams is an important option for the 
future linear $e^+e^-$ collider TESLA~\cite{tesla}, 
thus providing a direct measurement of 
event-shape observables in polarized $e^+e^-$ annihilation. 
\item NNLO predictions for  $(V+1)$-jet 
production at the RHIC polarized 
proton--proton collider and for $(2+1)$ jet production at a 
currently discussed polarized upgrade of the  
HERA collider do require the calculation of the two-loop  helicity amplitudes. 
These observables would then form part of a full NNLO determination of the 
polarized parton distribution functions in the proton.
\item The study of formal aspects of two-loop matrix elements, 
such as their 
collinear limits or their high energy behaviour can be carried out 
more elegantly on the basis of the underlying helicity amplitudes. 
\end{itemize}

Two-loop helicity amplitudes have up to now only been derived for 
$2\to 2$ bosonic scattering processes with all 
external legs on-shell: for 
$gg \to \gamma \gamma$~\cite{m5}, $\gamma \gamma \to 
\gamma \gamma$~\cite{m6,m10} and $gg \to gg$~\cite{m0,m9}. 
The latter calculation 
also confirmed earlier results for the squared two-loop 
 $gg \to gg$ matrix element~\cite{m4}. 
In the above calculations, which were all carried out within 
dimensional regularization~\cite{dreg1,dreg2,hv}
with $\dd=4-2\e$ space-time dimensions, 
two different methods were used to access the 
helicity structure of the matrix element: explicit contraction with 
the external polarization vectors~\cite{m5,m6,m0,m9} or projection onto 
the individual components of the Lorentz-invariant decomposition of the 
amplitude~\cite{m10}. Once these are applied to expose the helicity
structure, one is left with the task of computing a large number of 
two-loop integrals. Using integration-by-parts~\cite{hv,chet1,chet2} 
and Lorentz-invariance~\cite{gr} identities, these can be 
reduced~\cite{laporta} to 
a small number of so-called master integrals, which were derived 
for massless on-shell two-loop four-point 
functions
in~\cite{onshell1,onshell2,onshell3,onshell4,onshell5,onshell6}.
If an explicit contraction with the external polarization vectors is 
performed, one also has to compute two-loop integrals over the $(\dd-4)$
dimensional subspace of loop momenta, which reduce however to 
simple vacuum diagrams~\cite{m9}.
For $2\to 2$ scattering processes 
with external fermions  and all external legs on-shell ($e^+e^- \to e^+e^-$, $q\bar q \to q'\bar q'$, $q\bar q
\to q\bar q$, $q \bar q \to gg$, $q\bar q \to g \gamma$ and $q\bar q \to \gamma
\gamma$), only the 
squared, helicity-averaged two-loop matrix elements were computed  
so far~\cite{m1,m2,m3,m8}.

The method employed here to extract the two-loop helicity amplitudes for 
$e^+e^- \to q\bar q g$ is similar to the approach of~\cite{m10} by 
applying projections on all components of the Lorentz-invariant decomposition 
of the amplitude. Using this approach, the corresponding one-loop 
helicity amplitudes were derived in~\cite{gg}. 
The master integrals relevant in the present context are massless
four-point functions with three legs on-shell and one leg off-shell. The
complete set of these two-loop integrals was computed in~\cite{mi}, while 
earlier partial
results had been presented in~\cite{num,smirnov}\footnote{Note
that an alternative approach avoiding the need to use the integration-by-parts
and Lorentz-invariance identities to reduce the integrals appearing in the
Feynman diagrams to a basis set has recently been proposed\cite{muw,w2}. This
method relies on obtaining analytic expressions for the basic topologies with
arbitrary powers of the propagators and arbitrary dimensions, 
which can often be
found in terms of nested sums involving $\Gamma$-functions. 
The $\Gamma$-functions can
be directly expanded in $\epsilon$ and the nested sums related to
multiple polylogarithms.}.  
The master integrals in~\cite{mi} are expressed in terms of  two-dimensional
harmonic polylogarithms (2dHPLs).  The 2dHPLs are an extension of the harmonic 
polylogarithms (HPLs) of~\cite{hpl}. All HPLs and 2dHPLs that appear in the 
divergent parts of the planar master integrals have weight  $\leq 3$ and can
be  related to the more commonly known Nielsen  generalized 
polylogarithms~\cite{nielsen,bit} of suitable arguments.  The functions of
weight 4 appearing in the finite  parts of the master integrals can all be
represented, by their very  definition, as one-dimensional  integrals over
2dHPLs of weight 3, hence of Nielsen's  generalized
polylogarithms of suitable arguments according to the above  remark. A table
with all relations  is included in the appendix of~\cite{mi}. Numerical
routines providing an evaluation of   the HPLs~\cite{grnum1} and
2dHPLs~\cite{grnum2} are available. 

After carrying out 
ultraviolet renormalization of the amplitudes in the
$\overline{{\rm MS}}$ scheme, one is left with poles which are purely of 
infrared origin. The infrared pole structure of the amplitudes can be 
predicted using Catani's infrared factorization formula~\cite{catani}. 
We use this  formalism
to present the  infrared poles and the finite
parts  of the  helicity amplitudes 
 in a compact form. 

This paper is structured as follows: in Section~\ref{sec:method}, we outline
 the calculational method used to derive the helicity amplitudes 
and discuss the techniques used to extract the ultraviolet and infrared pole 
structure. We also elaborate on the relation to previous work. The 
two-loop helicity amplitudes are computed (in  the Weyl--van der Waerden
formalism, which is 
briefly described in the Appendix) 
in Section~\ref{sec:helicity}. Finally, Section~\ref{sec:conc}
contains a discussion of the results and conclusions.

\section{Method}
\setcounter{equation}{0}
\label{sec:method}

\subsection{Notation}
\label{subsec:defs}
We consider the production of a quark--antiquark--gluon 
system in electron--positron annihilation,
\begin{equation}
e^+(p_5) + e^-(p_6) \to \gamma^* (p_4) \longrightarrow q(p_1) + \bar q (p_2) + g(p_3)\; .
\end{equation}
It is convenient to define the invariants
\begin{equation}
\sab = (p_1+p_2)^2\;, \qquad \sac = (p_1+p_3)^2\;, \qquad 
\sbc = (p_2+p_3)^2\;,
\end{equation}
which fulfil
\begin{equation}
p_4^2  =(p_1+p_2+p_3)^2 = \sab + \sac + \sbc \equiv \sabc \; ,
\end{equation}
as well as the dimensionless invariants
\begin{equation}
x = \sab/\sabc\;, \qquad y = \sac/\sabc\;, \qquad z = \sbc/\sabc\;,
\end{equation}
which satisfy $x+y+z=1$.

The renormalized 
amplitude $|{\cal M}\rangle$ can be written as
\begin{equation}
\label{eq:Mdef}
|{\cal M}\rangle = V^\mu \S_\mu(q;g;\bar q)\; ,
\end{equation}
where $V^\mu$ represents the lepton current and $\S_\mu$ denotes the hadron current.
In a previous paper~\cite{3jme}, 
we have considered the unpolarized decay process
\begin{equation}
\gamma^* (p_4) \longrightarrow q(p_1) + \bar q (p_2) + g(p_3)\; .
\end{equation}
for which the amplitude is obtained from Eq.~(\ref{eq:Mdef}) by 
replacing the lepton current by the polarization vector of the virtual photon $\epsilon_4^\mu$.
The hadron current may be perturbatively decomposed as
\begin{equation}
\S_\mu(q;g;\bar q) = 
\sqrt{4\pi\alpha}e_q \sqrt{4\pi\alpha_s} \; {\bom
T}^{a}_{ij}\,\left(\S^{(0)}_\mu(q;g;\bar q) 
+ \left(\frac{\alpha_s}{2\pi}\right) \S^{(1)}_\mu(q;g;\bar q)
+ \left(\frac{\alpha_s}{2\pi}\right)^2 \S^{(2)}_\mu(q;g;\bar q) 
+ {\cal O}(\alpha_s^3)\right),
\label{eq:renorme}
\end{equation}
where $e_q$ denotes the quark charge,  
$a$ is the adjoint colour index for the gluon and $i$ and $j$ are 
the colour indices for quark and antiquark.
$\alpha_s$ is the QCD coupling constant at the renormalization scale $\mu$, 
and the $\S^{(i)}_\mu$ are the $i$-loop contributions to the 
renormalized amplitude. Renormalization of ultraviolet divergences is 
performed in the $\overline{{\rm MS}}$ scheme.

\subsection{The general tensor}
\label{subsec:genten}
The most general tensor structure for the hadron current $\S_\mu(q;g;\bar q)$ is
\begin{eqnarray}
\S_\mu(q;g;\bar q) &=&  
\bar u(p_1)\slash \!\!\! p_3 u(p_2) \left(
  A_{11} \epsilon_3.p_1~p_{1\mu} 
+ A_{12} \epsilon_3.p_1~p_{2\mu}
+ A_{13} \epsilon_3.p_1~p_{3\mu}
\right)
\nonumber \\
&+& \bar u(p_1)\slash \!\!\!p_3 u(p_2) \left(
  A_{21} \epsilon_3.p_2~p_{1\mu}
+ A_{22}  \epsilon_3.p_2 ~p_{2\mu}
+ A_{23}  \epsilon_3.p_2 ~p_{3\mu}
\right)
\nonumber \\
&+& \bar u(p_1)\gamma_\mu u(p_2) \left(
  B_{1} \epsilon_3.p_1
+ B_{2} \epsilon_3.p_2
\right)
\nonumber \\
&+&\bar u(p_1)\slash \!\!\!\epsilon_3 u(p_2) \left( 
 C_{1} p_{1\mu}
+C_{2} p_{2\mu}
+C_{3} p_{3\mu}
\right)
\nonumber \\
&+& D_{1} \bar u(p_1)\slash \!\!\!\epsilon_3 \slash \!\!\!p_3 \gamma_\mu u(p_2) 
\nonumber \\
&+& D_{2} \bar u(p_1)\gamma_\mu \slash \!\!\!p_3 \slash \!\!\!\epsilon_3 u(p_2) \; ,
\end{eqnarray}
where the constraint $\epsilon_3 \cdot p_3 = 0$ has been applied.
All coefficients are functions of $\sac$, $\sbc$ and $\sabc$.
The above tensor structure is a priori $\dd$-dimensional, since the 
dimensionality of the external states has not yet been specified. 
The hadron current is conserved and
satisfies
\begin{equation}
\S_\mu(q;g;\bar q)~p_{4}^{\mu} = 0\; ;
\end{equation}
it must also obey the QCD Ward identity when the gluon polarization vector $\epsilon_3$ is
replaced with the gluon momentum,
\begin{equation}
\S_\mu(q;g;\bar q) (\epsilon_3 \to p_3) = 0.
\end{equation}
These constraints yield relations amongst the 13 distinct tensor structures and 
applying
these identities gives the gauge-invariant form of the tensor,
\begin{eqnarray}
\S_\mu(q;g;\bar q)~~=&& A_{11}(\sac,\sbc,\sabc) T_{11\mu} 
                     + A_{12}(\sac,\sbc,\sabc) T_{12\mu} 
                     + A_{13}(\sac,\sbc,\sabc) T_{13\mu} \nonumber \\
&+&
A_{21}(\sac,\sbc,\sabc)  T_{21\mu}
+ A_{22}(\sac,\sbc,\sabc)  T_{22\mu}
+A_{23}(\sac,\sbc,\sabc)  T_{23\mu}\nonumber \\
&+&B(\sac,\sbc,\sabc)  T_{\mu},
\end{eqnarray}
where $A_{ij}$ and $B$ are gauge-independent functions and 
the tensor structures $T_{IJ\mu}$ and $T_{\mu}$ are given by
\begin{eqnarray}
T_{1J\mu} &=&\bar u(p_1)\slash \!\!\! p_3 u(p_2)
\epsilon_3.p_1p_{J\mu} 
-\frac{s_{13}}{2}
\bar u(p_1)\slash \!\!\! \epsilon_3 u(p_2)p_{J\mu}
+\frac{s_{J4}}{4} 
\bar u(p_1)\slash \!\!\!\epsilon_3 \slash \!\!\!p_3 \gamma_\mu
u(p_2),\\
T_{2J\mu} &=&\bar u(p_1)\slash \!\!\! p_3 u(p_2)
\epsilon_3.p_2p_{J\mu} 
-\frac{s_{23}}{2}
\bar u(p_1)\slash \!\!\! \epsilon_3 u(p_2)p_{J\mu}
+\frac{s_{J4}}{4} 
\bar u(p_1)\gamma_\mu \slash \!\!\!p_3 \slash \!\!\!\epsilon_3
u(p_2),\\
T_{\mu}&=& s_{23}\left(
\bar u(p_1)\gamma_\mu u(p_2)  
\epsilon_3.p_1
+\frac{1}{2}
\bar u(p_1)\slash \!\!\!\epsilon_3 \slash \!\!\!p_3 \gamma_\mu
u(p_2) \right)
\nonumber \\
&-&s_{13}\left(
\bar u(p_1)\gamma_\mu u(p_2)  
\epsilon_3.p_2
+\frac{1}{2}
\bar u(p_1)\gamma_\mu \slash \!\!\!p_3 \slash \!\!\!\epsilon_3
u(p_2) \right).
\end{eqnarray}
Each of the tensor structures satisfies 
both current conservation and the QCD Ward identity.
The coefficients are 
further related by symmetry under the interchange of the quark and antiquark,
\begin{eqnarray}
A_{21}(\sac,\sbc,\sabc) &=& - A_{12}(\sbc,\sac,\sabc),\nonumber \\
A_{22}(\sac,\sbc,\sabc) &=& - A_{11}(\sbc,\sac,\sabc),\nonumber \\
A_{23}(\sac,\sbc,\sabc) &=& - A_{13}(\sbc,\sac,\sabc),\nonumber \\
B(\sac,\sbc,\sabc) &=& B(\sbc,\sac,\sabc).
\end{eqnarray}

\subsection{Projectors for the tensor coefficients}
\label{subsec:projectors}

The coefficients $A_{IJ}$ and $B$  may be easily extracted from a Feynman diagram calculation,
using projectors such that
\begin{equation}
\sum_{\rm spins} {\cal P}(X) ~\epsilon_4^\mu  \S_\mu(q;g;\bar q) = X(\sac,\sbc,\sabc).
\end{equation}
The explicit forms for the seven projectors in $d$ space-time dimensions are,
\begin{eqnarray}
{\cal P}(A_{11}) &=& 
\frac{(\sbc\sabc \dd+\sac\sab \dmb)}{2\sac^3\sab^2\dmc\sabc}\Taa
-\frac{(\sac+\sbc) \dmb}{2\sac^2\sab^2\dmc\sabc}\Tab\nonumber \\
&& 
-\frac{((\sbc +\sab)\dd+2\sac)}{2\sab\sac^3\dmc\sabc}\Tac
-\frac{(\sbc\sabc\dmb+\sac\sab \dmd)}{2\sbc\sab^2\sac^2\sabc\dmc}\Tba\nonumber \\
&&
+\frac{(\sac+\sbc)\dmd}{2\dmc\sab^2\sabc\sac\sbc}\Tbb
+\frac{(\sbc+\sab)\dmd}{2\sbc\sab\sac^2\sabc\dmc}\Tbc\nonumber \\
&&
-\frac{1}{2\sac^2\sab^2\dmc}\T,\nonumber \\
{\cal P}(A_{12}) &=& 
-\frac{(\sac+\sbc) \dmb}{2\sac^2\sab^2\dmc\sabc}\Taa
+\frac{\dmb(\sbc\sab \dmd+\sac\sabc \dmb)}{2\sac^2\sab^2\sbc\dmc\sabc\dmd}\Tab\nonumber \\
&&
-\frac{\dmb(\sac+\sab)}{2\sac^2\sab\sbc\dmc\sabc}\Tac
+\frac{(\dmf \dmb(\sac+\sbc)-4\sab)}{2\dmd\sab^2\sac\sbc\sabc\dmc}\Tba \nonumber \\
&&
-\frac{(\sbc\sab \dmd+\sac\sabc \dmb)}{2\sab^2\sac\sbc^2\dmc\sabc}\Tbb
+\frac{(2\sbc+(\sac +\sab) \dmb)}{2\sab\sac\sbc^2\sabc\dmc}\Tbc\nonumber \\
&&
-\frac{\dmb}{2\dmd\sab^2\sac\sbc\dmc}\T,\nonumber \\
{\cal P}(A_{13}) &=& 
-\frac{((\sbc +\sab)\dd+2\sac)}{2\sab\sac^3\dmc\sabc}\Taa
-\frac{ \dmb(\sac+\sab)}{2\sac^2\sab\sbc\dmc\sabc}\Tab\nonumber \\
&&
+\frac{(\sac\sbc \dmb+\sab\sabc\dd)}{2\sac^3\sab\sbc\sabc\dmc}\Tac
+\frac{((\sab+\sbc) \dmb+2\sac)}{2\sac^2\sab\sbc\dmc\sabc}\Tba\nonumber \\
&&
+\frac{(\sac+\sab)\dmd}{2\sab\sac\sbc^2\sabc\dmc}\Tbb
-\frac{(\sac+\sab)(\sbc+\sab)\dmd}{2\sac^2\sab\sbc^2\sabc\dmc}\Tbc\nonumber \\
&&
+\frac{1}{2\sbc\sac^2\sab\dmc}\T,\nonumber \\
{\cal P}(A_{21}) &=& 
-\frac{(\sbc\sabc \dmb+\sac\sab \dmd)}{2\sac^2\sab^2\sbc\dmc\sabc}\Taa
+\frac{(-4\sab+(\sac+\sbc)\dmf \dmb)}{2\dmd\sab^2\sac\sbc\sabc\dmc}\Tab\nonumber \\
&&
+\frac{(\sbc+\sab) \dmb+2\sac)}{2\sac^2\sab\sbc\dmc\sabc}\Tac
+\frac{\dmb(\sbc\sabc\dmb+\sac\sab \dmd)}{2\sab^2\sac\sbc^2\sabc\dmc\dmd}\Tba\nonumber \\
&&
-\frac{(\sac+\sbc) \dmb}{2\sab^2\sbc^2\sabc\dmc}\Tbb
-\frac{(\sbc+\sab) \dmb}{2\sac\sbc^2\sab\dmc\sabc}\Tbc\nonumber \\
&&
+\frac{\dmb}{2\dmd\sab^2\sac\sbc\dmc}\T,\nonumber \\
{\cal P}(A_{22}) &=& \frac{(\sac+\sbc)\dmd}{2\sac\sab^2\sbc\dmc\sabc}\Taa
-\frac{(\sbc\sab \dmd+\sac\sabc \dmb)}{2\sab^2\sac\sbc^2\dmc\sabc}\Tab\nonumber \\
&&
+\frac{(\sac+\sab)\dmd}{2\sac\sbc^2\sab\dmc\sabc}\Tac
-\frac{(\sac+\sbc) \dmb}{2\sab^2\sbc^2\sabc\dmc}\Tba\nonumber \\
&&
+\frac{(\sbc\sab \dmb+\sac\sabc\dd)}{2\sbc^3\sab^2\dmc\sabc}\Tbb
-\frac{(\sac\dd+\sab\dd+2\sbc)}{2\sab\sbc^3\sabc\dmc}\Tbc\nonumber \\
&&
+\frac{1}{2\sbc^2\sab^2\dmc}\T,\nonumber \\
{\cal P}(A_{23}) &=& \frac{(\sbc+\sab)\dmd}{2\sbc\sab\sac^2\sabc\dmc}\Taa
+\frac{(2\sbc+(\sac+\sab) \dmb)}{2\sac\sbc^2\sab\dmc\sabc}\Tab\nonumber \\
&&
-\frac{(\sac+\sab)(\sbc+\sab)\dmd}{2\sac^2\sab\sbc^2\sabc\dmc}\Tac
-\frac{(\sbc+\sab) \dmb}{2\sac\sbc^2\sab\dmc\sabc}\Tba\nonumber \\
&&
-\frac{((\sac +\sab)\dd+2\sbc)}{2\sab\sbc^3\sabc\dmc}\Tbb
+\frac{(\sac\sbc \dmb+\sab\sabc\dd)}{2\sac\sab\sbc^3\sabc\dmc}\Tbc\nonumber \\
&&
-\frac{1}{2\sbc^2\sac\sab\dmc}\T,\nonumber \\
{\cal P}(B) &=& -\frac{1}{2\dmc\sab^2\sac^2}\Taa
-\frac{\dmb}{2\dmd\sab^2\sac\sbc\dmc}\Tab
+\frac{1}{2\sbc\sab\dmc\sac^2}\Tac\nonumber \\
&&
+\frac{ \dmb}{2\dmd\sab^2\sac\sbc\dmc}\Tba
+\frac{1}{2\sbc^2\sab^2\dmc}\Tbb
-\frac{1}{2\sac\sab\dmc\sbc^2}\Tbc\nonumber \\
&&
+\frac{1}{2\dmd\sab^2\sac\sbc}\T.
\end{eqnarray} 

\subsection{The perturbative expansion of the tensor coefficients}
\label{subsec:perturbative}

Each of the unrenormalized
coefficients $A_{IJ}$ and $B$ has a perturbative expansion of the form
\begin{eqnarray}
A_{IJ}^{{\rm un}} &=&  \sqrt{4\pi\alpha}e_q \sqrt{4\pi\alpha_s} \; {\bom
T}^{a}_{ij}\,\left[
A_{IJ}^{(0),{\rm un}}  
+ \left(\frac{\alpha_s}{2\pi}\right) A_{IJ}^{(1),{\rm un}}  
+ \left(\frac{\alpha_s}{2\pi}\right)^2 A_{IJ}^{(2),{\rm un}} 
+ {\cal O}(\alpha_s^3) \right] \;,\nonumber \\
B^{{\rm un}} &=&  \sqrt{4\pi\alpha}e_q \sqrt{4\pi\alpha_s} \; {\bom T}^{a}_{ij}\,\left[
B^{(0),{\rm un}}  
+ \left(\frac{\alpha_s}{2\pi}\right) B^{(1),{\rm un}}  
+ \left(\frac{\alpha_s}{2\pi}\right)^2 B^{(2),{\rm un}} 
+ {\cal O}(\alpha_s^3) \right] \;,\nonumber \\
\end{eqnarray}
where the dependence on $(\sac,\sbc,\sabc)$ is implicit.
At tree level,
\begin{eqnarray}
A_{IJ}^{(0),{\rm un}}(\sac,\sbc,\sabc) &=& 0,\\
B^{(0),{\rm un}}(\sac,\sbc,\sabc) 
 &=& \frac{2}{s_{13}s_{23}} \;.
\end{eqnarray}
The one-loop contributions can be written in terms of the one-loop box
integral in
$\dd=6-2\epsilon$
 dimensions, Box$^6(s_{ij},s_{ik},s_{ijk})$,
 and the one-loop bubble, Bub$(s_{ij})$, as follows:
\begin{eqnarray}
\lefteqn{A_{11}^{(1),{\rm un}}(\sac,\sbc,\sabc) =}\nonumber \\ 
&& N \Biggl (
  -\frac{\dmd}{2(\sac+\sab)\sac} ~\Babc 
  -\frac{\dmd}{2\sab\sac} ~\bac 
  \nonumber \\ &&
  \hspace{1cm}-\frac{(\dmb\sbc\sab+\dmd\sbc\sac+4\sab(\sab+\sac))}{2\sab\sac(\sac+\sab)^2} ~\bbc 
  \nonumber \\ &&
  \hspace{1cm}-\frac{\dmd(4\sab+\dmb\sbc)}{4\sab\sac} ~\Boxyz 
\Biggr)\nonumber \\ 
&&
+\frac{1}{N}\Biggl (
  \frac{\dmd}{2(\sac+\sab)\sac} ~\Babc 
  +\frac{\dd}{2\sac^2} ~\bab 
  \nonumber \\ &&
  \hspace{1cm}+\frac{(\sab+\sac)(\dd\sbc+4\sac)+2\sbc\sac}{2(\sac+\sab)^2\sac^2} ~\bbc 
  \nonumber \\ &&
  \hspace{1cm}+\frac{\dmd\dmf}{4\sac} ~\Boxxy 
+\frac{\dmb(\dd\sbc+4\sac)}{4\sac^2} ~\Boxxz 
\Biggr),
\label{eq:aNLO}\\
\lefteqn{A_{12}^{(1),{\rm un}}(\sac,\sbc,\sabc) =}\nonumber \\ 
&& 
N \Biggl (
  -\frac{\dmj}{2\sab(\sbc+\sab)} ~\bac 
-\frac{(\dmj\sac-4\sab)}{2\sab\sac(\sac+\sab)} ~\bbc 
  \nonumber \\ &&
  \hspace{1cm}+\frac{(4\dmd\sab-\dmb\dmj\sac)}{4\sab\sac} ~\Boxyz 
\Biggr)\nonumber \\ 
&&
+\frac{1}{N}\Biggl (
  \frac{\dmb}{2\sbc\sac} ~\bab 
+\frac{(\dmb\sab+2\dmf\sbc)}{2\sbc\sab(\sbc+\sab)} ~\bac 
  \nonumber \\ &&
  \hspace{1cm}+\frac{\dmf(\sab+2\sac)}{2\sab\sac(\sac+\sab)} ~\bbc 
  \nonumber \\ &&
  \hspace{1cm}+\frac{(\dmb^2\sab\sac+4\dmd\sab\sbc+2\dmd\dmf\sac\sbc)}{4\sab\sac\sbc} ~\Boxxy 
  \nonumber \\ &&
  \hspace{1cm}+\frac{\dmf(\dmb\sab+2\dmd\sac)}{4\sab\sac} ~\Boxxz 
\Biggr),\\
\lefteqn{A_{13}^{(1),{\rm un}}(\sac,\sbc,\sabc) =}\nonumber \\ 
&& 
N \Biggl (
  \frac{\dmf}{2(\sac+\sab)\sac} ~\bbc 
  +\frac{\dmd\dmf}{4\sac} ~\Boxyz 
\Biggr)\nonumber \\ 
&&
+\frac{1}{N}\Biggl (
  -\frac{\dmd}{\sac(\sbc+\sac)} ~\Babc 
  -\frac{\dmd}{2\sbc\sac} ~\bac 
\nonumber \\ &&
 \hspace{1cm}+\frac{(4\sab\sac^2-\dd\sab(\sac+\sbc)^2-2\dmb\sac\sbc(\sac+\sbc))}{2(\sbc+\sac)^2\sac^2\sbc} ~\bab 
\nonumber \\ &&
  \hspace{1cm}-\frac{(2\dmc\sac+\dd\sab)}{2\sac^2(\sac+\sab)} ~\bbc 
-\frac{\dmd(\dmb\sab+4\sbc)}{4\sbc\sac} ~\Boxxy 
\nonumber \\ &&
  \hspace{1cm}-\frac{\dmb(\dd\sab+2\dmd\sac)}{4\sac^2} ~\Boxxz 
\Biggr), \\
\lefteqn{B^{(1),{\rm un}}(\sac,\sbc,\sabc) =}\nonumber \\ 
&& 
N \Biggl (
  \frac{\dd^2-3\dd+4}{4\dmd\sac\sbc} ~\Babc 
\nonumber \\ &&
 \hspace{1cm}+\frac{(4\dmc\sab(\sab+\sbc)+\dmd(\dd-7)\sbc\sac)}{2\sab\sbc(\sbc+\sab)\sac\dmd}
~\bac 
\nonumber \\ &&
  \hspace{1cm}+\frac{(4\dmc\sab^2+\dmb(\dd-7)\sac\sbc)}{8\sab\sac\sbc} ~\Boxyz 
\Biggr)\nonumber \\ 
&&
+\frac{1}{N}\Biggl (
  \frac{(7\dd-16-\dd^2)}{4\dmd\sac\sbc} ~\Babc  
  +\frac{(16-5\dd)}{4\dmd\sac\sbc} ~\bab \nonumber \\ && 
  \hspace{1cm}-\frac{(\sab+\dmf\sbc)}{2(\sbc+\sab)\sbc\sab} ~\bac \nonumber \\ && 
  \hspace{1cm}-\frac{(4\dmc\sbc\sab+\dmd\dmf\sac\sbc+\dmb\sab\sac)}{4\sab\sac\sbc} ~\Boxxy
  \Biggr) 
  \nonumber \\ && 
  + \Biggl \{ \sac \leftrightarrow \sbc \Biggr \}.
\label{eq:bNLO}
\end{eqnarray}
Explicit expansions of the one-loop integrals 
around $\epsilon \sim 0$ in terms of HPLs
and 2dHPLs are listed in Appendix~A of~\cite{3jme}. 

Similarly, the unrenormalized two-loop $A_{IJ}^{(2),{\rm un}}$ and $B^{(2),{\rm un}}$ 
coefficients were obtained analytically 
(making extensive use of the computer algebra programs MAPLE~\cite{maple},
FORM2~\cite{form2} and FORM3~\cite{form3}, where the latter two 
are particularly well suited for handling the large-size expressions arising 
at intermediate stages of the calculation) in
terms of  a basis set of two-loop master integrals.  This basis set 
comprises  14 planar
topologies and 5 non-planar topologies. Five of the topologies require more than one
master integral, so that in total 24 master integrals are needed. 
A more detailed
discussion can be found in  Ref.~\cite{3jme}.  However, we note that Laurent
expansions for each of these master integrals have
 been derived in~\cite{mi} by solving
differential equations for the master integrals (equations that are differential with
respect to the momentum scales involved in the diagram). 
The $\e$-expansions of 
$A_{IJ}^{(2),{\rm un}}$ and $B^{(2),{\rm un}}$ can therefore be obtained by directly substituting the
$\e$-expansions of the individual master integrals.

\subsection{Relation to previous work}
\label{subsec:earlier}

We have considered the case where the correlations with the 
lepton current are ignored in a
previous paper~\cite{3jme}.
In this instance,  the squared amplitude for the process $\gamma^* \to q\bar q
g$, summed over spins, colours and quark flavours, 
was denoted by
\begin{equation}
\langle{\cal M}|{\cal M}\rangle = \sum | \epsilon_4 \cdot \S(q;g;\bar q)  | ^2
= {\cal T} (x,y,z)\; .
\end{equation}
The perturbative expansion of ${\cal T} (x,y,z)$ at renormalization scale 
$\mu^2 = q^2 = s_{123}$ reads:
\begin{eqnarray}
{\cal T} (x,y,z) &=& 16\pi^2\alpha\sum_q e_q^2 \alpha_s(q^2)\Bigg[
{\cal T}^{(2)} (x,y,z) + 
\left(\frac{\alpha_s(q^2)}{2\pi}\right){\cal T}^{(4)} (x,y,z) \nonumber \\
&& \hspace{1.3cm}
+ \left(\frac{\alpha_s(q^2)}{2\pi}\right)^2{\cal T}^{(6)} (x,y,z) 
 + {\cal O}(\alpha_s^3(q^2)) \Bigg] \;,
\end{eqnarray}
where 
\begin{eqnarray}
\label{eq:T2}
{\cal T}^{(2)} (x,y,z) &=& \langle{\cal M}^{(0)}|{\cal M}^{(0)}\rangle 
= 4 V (1-\e)\left[ (1-\e)\left(\frac{y}{z}+\frac{z}{y}\right) 
+\frac{2(1-y-z)-2\e yz}{yz}\right]\;,\\
\label{eq:T4}
{\cal T}^{(4)} (x,y,z) &=& 
\langle{\cal M}^{(0)}|{\cal M}^{(1)}\rangle +
\langle{\cal M}^{(1)}|{\cal M}^{(0)}\rangle \; ,\\
\label{eq:T6}
{\cal T}^{(6)} (x,y,z) &=& 
\langle{\cal M}^{(1)}|{\cal M}^{(1)}\rangle +
\langle{\cal M}^{(0)}|{\cal M}^{(2)}\rangle +
\langle{\cal M}^{(2)}|{\cal M}^{(0)}\rangle \;,
\end{eqnarray}
where $V=N^2-1$, with $N$ the number of colours. 
${\cal T}^{(4)} (x,y,z)$ was first derived in~\cite{ert1,ert2} through to ${\cal
O}(\epsilon^0)$ while an explicit expression for it to all orders in $\epsilon$ 
was given in~\cite{3jme}.
The contribution to ${\cal T}^{(6)} (x,y,z)$ from the interference 
of two-loop and tree diagrams
\begin{equation}
{\cal T}^{(6,[2\times 0])} (x,y,z) = 
\langle{\cal M}^{(0)}|{\cal M}^{(2)}\rangle +
\langle{\cal M}^{(2)}|{\cal M}^{(0)}\rangle \;,
\end{equation}
as well as the one-loop self-interference 
\begin{equation}
{\cal T}^{(6,[1\times 1])} (x,y,z) = 
\langle{\cal M}^{(1)}|{\cal M}^{(1)}\rangle \;
\end{equation}
were first derived in~\cite{3jme}.

It is straightforward to obtain the interference of the tree and $i$-loop
amplitudes in terms of the tensor coefficients, $A_{IJ}$ and $B$.  We find
\begin{eqnarray}
\label{eq:square}
\lefteqn{
\langle{\cal M}^{(0)}|{\cal M}^{(i)}\rangle =}\nonumber \\ &&
\frac{V}{2}\, \Biggl\{ 
2 (1-\epsilon) \left((\sab \sabc+\sab \sac+\sac \sbc)-\epsilon(\sac+\sbc)(\sab+\sac)\right) A_{11}^{(i)} (\sac,\sbc,\sabc) 
\nonumber \\
&&\phantom{\biggr\{}+\left(2(\sab+\sbc)^2-2\epsilon \left(\sabc \sbc +(\sab+\sbc)^2\right) + 2\epsilon^2
(\sac+\sbc)(\sab+\sbc) \right) A_{12}^{(i)}(\sac,\sbc,\sabc)\nonumber \\
&&\phantom{\biggr\{}
+2\left(\sbc-\epsilon(\sac+\sbc))(\sabc-\epsilon(\sac+\sbc)\right)A_{13}^{(i)}(\sac,\sbc,\sabc)\nonumber \\
&&\phantom{\biggr\{}+2\left(\sac^2+\sbc^2+2\sab\sabc-2\epsilon(\sabc^2-\sab\sac-\sab\sbc-\sac\sbc)+\epsilon^2(\sac+\sbc)^2\right) B^{(i)}(\sac,\sbc,\sabc)
\nonumber \\
&&\hspace{2cm}+ \{p_1 \leftrightarrow p_2 \}
\Biggr\}.
\end{eqnarray}
The above relation holds for the unrenormalized as well as for 
the renormalized matrix element, involving the appropriate unrenormalized or 
renormalized tensor coefficients respectively.
Similar, but more lengthy, expressions can easily be obtained for the
interference of $i$- and $j$-loop amplitudes.
We have checked that inserting the expressions for $A_{IJ}^{(i)}$ 
and $B^{(i)}$ into
Eq.~(\ref{eq:square}) reproduces our earlier results~\cite{3jme} at the 
one- and two-loop level both at the master integral level and after making an expansion in
$\epsilon$.

\subsection{Ultraviolet renormalization}
\label{subsec:renorm}

The renormalization of the matrix element is carried out by replacing 
the bare coupling $\alpha_0$ with the renormalized coupling 
$\alpha_s\equiv \alpha_s(\mu^2)$,
evaluated at the renormalization scale $\mu^2$
\begin{equation}
\alpha_0\mu_0^{2\e} S_\e = \alpha_s \mu^{2\e}\left[
1- \frac{\beta_0}{\e}\left(\frac{\alpha_s}{2\pi}\right) 
+\left(\frac{\beta_0^2}{\e^2}-\frac{\beta_1}{2\e}\right)
\left(\frac{\alpha_s}{2\pi}\right)^2+{\cal O}(\alpha_s^3) \right]\; ,
\end{equation}
where
\begin{displaymath}
S_\e =(4\pi)^\e e^{-\e\gamma}\qquad \mbox{with Euler constant }
\gamma = 0.5772\ldots
\end{displaymath}
and $\mu_0^2$ is the mass parameter introduced 
in dimensional regularization~\cite{dreg1,dreg2,hv} to maintain a 
dimensionless coupling 
in the bare QCD Lagrangian density; $\beta_0$ and $\beta_1$ are the first 
two coefficients of the QCD $\beta$-function:
\begin{equation}
\beta_0 = \frac{11 \CA - 4 T_R \NF}{6},  \qquad 
\beta_1 = \frac{17 \CA^2 - 10 C_A T_R \NF- 6C_F T_R \NF}{6}\;,
\end{equation}
with the QCD colour factors
\begin{equation}
\CA = N,\qquad C_F = \frac{N^2-1}{2N},
\qquad T_R = \frac{1}{2}\; .
\end{equation}

We denote the $i$-loop contribution to the unrenormalized coefficients by 
$A_{IJ}^{(i),{\rm un}}$ and $B^{(i),{\rm un}}$, using the same normalization as 
for the decomposition of the renormalized amplitude (\ref{eq:renorme}); the 
dependence on $(\sac,\sbc,\sabc)$ is always understood implicitly. 
The renormalized coefficients are then obtained as
\begin{eqnarray}
A_{IJ}^{(0)}  &=&0,
 \nonumber \\
A_{IJ}^{(1)}  &=& 
S_\e^{-1} A_{IJ}^{(1),{\rm un}} 
 ,  \nonumber \\
A_{IJ}^{(2)} &=& 
S_\e^{-2} A_{IJ}^{(2),{\rm un}}  
-\frac{3\beta_0}{2\e} S_\e^{-1}
A_{IJ}^{(1),{\rm un}}  ,
\end{eqnarray}
and
\begin{eqnarray}
B^{(0)}  &=& B^{(0),{\rm un}} ,
 \nonumber \\
B^{(1)}  &=& 
S_\e^{-1} B^{(1),{\rm un}} 
-\frac{\beta_0}{2\e} B^{(0),{\rm un}}  ,  \nonumber \\
B^{(2)} &=& 
S_\e^{-2} B^{(2),{\rm un}}  
-\frac{3\beta_0}{2\e} S_\e^{-1}
B^{(1),{\rm un}}  
-\left(\frac{\beta_1}{4\e}-\frac{3\beta_0^2}{8\e^2}\right)
B^{(0),{\rm un}}.
\end{eqnarray}

For the remainder of this paper we will set the renormalization scale
$\mu^2 = q^2$. 
The full scale dependence of the tensor coefficients is given by
\begin{eqnarray}
A_{IJ} &=& \sqrt{4\pi\alpha}e_q \sqrt{4\pi\alpha_s} \; {\bom
T}^{a}_{ij}\, \bigg\{
\left(\frac{\alpha_s(\mu^2)}{2\pi}\right)
A_{IJ}^{(1)}  
+ \left(\frac{\alpha_s(\mu^2)}{2\pi}\right)^2
\bigg[A_{IJ}^{(2)} 
+\frac{3\beta_0}{2}A_{IJ}^{(1)}  
\ln\left({\mu^2\over q^2}\right)  
\bigg]
 + {\cal O}(\alpha_s^3) \bigg\},\nonumber \\
B &=& \sqrt{4\pi\alpha}e_q \sqrt{4\pi\alpha_s} \; {\bom
T}^{a}_{ij}\, \bigg\{
B^{(0)}  + \left(\frac{\alpha_s(\mu^2)}{2\pi}\right)
\left[
B^{(1)}  
+\frac{\beta_0}{2} B^{(0)} \ln\left({\mu^2\over q^2}\right)
\right]
\nonumber \\
&&+ \left(\frac{\alpha_s(\mu^2)}{2\pi}\right)^2
\bigg[B^{(2)} 
+\biggl(\frac{3\beta_0}{2}B^{(1)} +\frac{\beta_1}{2}B^{(0)} \biggr) 
\ln\left({\mu^2\over q^2}\right)  
+\frac{3 \beta_0^2}{8} B^{(0)}  
\ln^2\left({\mu^2\over q^2}\right)\bigg]
 + {\cal O}(\alpha_s^3) \bigg\}.
\end{eqnarray}

\subsection{Infrared behaviour of the tensor coefficients}
\label{subsec:infrared}

After performing ultraviolet renormalization,
the amplitudes still
contain singularities, which are of infrared origin and will be  analytically
cancelled by those occurring in radiative processes of the
same order.
Catani~\cite{catani} has shown how to organize the 
infrared pole structure of the one- and two-loop contributions renormalized in the 
\MSbar\ scheme in terms of the tree and renormalized one-loop amplitudes.
The same procedure applies to the tensor coefficients.
In particular, the infrared behaviour of the one-loop coefficients is given by
\begin{eqnarray}
A_{IJ}^{(1)} &=& A_{IJ}^{(1),{\rm finite}},\nonumber \\
B^{(1)} &=& {\bom I}^{(1)}(\epsilon) B^{(0)} + B^{(1),{\rm finite}},
\end{eqnarray}
while the two-loop singularity structure is
\begin{eqnarray}
A_{IJ}^{(2)} &=& {\bom I}^{(1)}(\epsilon) A_{IJ}^{(1)}+A_{IJ}^{(2),{\rm finite}},\nonumber \\
B^{(2)} &=& \Biggl (-\frac{1}{2}  {\bom I}^{(1)}(\epsilon) {\bom I}^{(1)}(\epsilon)
-\frac{\beta_0}{\epsilon} {\bom I}^{(1)}(\epsilon) 
+e^{-\epsilon \gamma } \frac{ \Gamma(1-2\epsilon)}{\Gamma(1-\epsilon)} 
\left(\frac{\beta_0}{\epsilon} + K\right)
{\bom I}^{(1)}(2\epsilon) + {\bom H}^{(2)}(\epsilon) 
\Biggr )B^{(0)}\nonumber \\
&& + {\bom I}^{(1)}(\epsilon) B^{(1)}+ B^{(2),{\rm finite}},
\label{eq:polesa}
\end{eqnarray}
where the constant $K$ is
\begin{equation}
K = \left( \frac{67}{18} - \frac{\pi^2}{6} \right) \CA - 
\frac{10}{9} T_R \NF.
\end{equation}
The finite remainders $A_{IJ}^{(i),{\rm finite}}$ and $B^{(i),{\rm finite}}$ remain to be
calculated.

For this particular process, there is only one colour structure present at
tree level which, in terms of the gluon colour $a$ and the quark
and antiquark
colours $i$ and $j$, is simply $\bom{T}^{a}_{ij}$. Adding higher loops does not
introduce additional colour structures, and the amplitudes are therefore
vectors in a one-dimensional space.  Similarly, 
the infrared singularity operator $\bom{I}^{(1)}(\epsilon)$ is a $1 \times 1$ matrix in the colour space
and is given by
\begin{equation}
\bom{I}^{(1)}(\epsilon)
=
- \frac{e^{\epsilon\gamma}}{2\Gamma(1-\epsilon)} \Biggl[
N \left(\frac{1}{\epsilon^2}+\frac{3}{4\epsilon}+\frac{\beta_0}{2N\epsilon}\right) 
\left({\tt S}_{13}+{\tt S}_{23}\right)-\frac{1}{N}
\left(\frac{1}{\epsilon^2}+\frac{3}{2\epsilon}\right)
{\tt S}_{12}\Biggr ]\; ,\label{eq:I1}
\end{equation}
where (since we have set $\mu^2 = s_{123}$)
\begin{equation}
{\tt S}_{ij} = \left(-\frac{s_{123}}{s_{ij}}\right)^{\epsilon}.
\end{equation}
Note that on expanding ${\tt S}_{ij}$,
imaginary parts are generated, the sign of which is fixed by the small imaginary
part $+i0$ of $s_{ij}$.
The origin of the various terms in Eq.~(\ref{eq:I1}) is straightforward.  Each parton pair $ij$
in the event forms a radiating antenna of scale $s_{ij}$.  
Terms proportional to ${\tt S}_{ij}$ are cancelled by real radiation emitted from leg
$i$ and absorbed by leg $j$. The soft singularities ${\cal O}(1/\epsilon^2)$ are independent of
the identity of the participating partons and are universal.
However, the collinear singularities depend on the identities of the participating partons.  

Finally, the term of Eq.~(\ref{eq:polesa}) that involves 
${\bom H}^{(2)}(\epsilon)$ 
produces only a single pole in $\epsilon$ and is given by 
\begin{equation}
\label{eq:htwo}
{\bom H}^{(2)}(\epsilon)
=\frac{e^{\epsilon \gamma}}{4\,\epsilon\,\Gamma(1-\epsilon)} H^{(2)} \;,  
\end{equation}
where the constant $H^{(2)}$ is renormalization-scheme-dependent.
As with the single-pole parts of $\bom{I}^{(1)}(\epsilon)$,
the process-dependent
$H^{(2)}$ can be constructed by counting the number of
radiating partons present in the event.
In our case, there is a quark--antiquark pair and a gluon present in the final
state, so that 
\begin{equation}
H^{(2)} =  2H^{(2)}_{q}+H^{(2)}_g\; ,
\end{equation}
where, in the \MSbar\ scheme:
\begin{eqnarray}
H^{(2)}_g &=&  
\left(\frac{1}{2}\zeta_3+{\frac {5}{12}}+ {\frac {11\pi^2}{144}}
\right)N^2
+{\frac {5}{27}}\,\NF^2
+\left (-{\frac {{\pi }^{2}}{72}}-{\frac {89}{108}}\right ) N \NF 
-\frac{\NF}{4N}, \\
H^{(2)}_q &=&
\left({7\over 4}\zeta_3+{\frac {409}{864}}- {\frac {11\pi^2}{96}}
\right)N^2
+\left(-{1\over 4}\zeta_3-{41\over 108}-{\pi^2\over 96}\right)
+\left(-{3\over 2}\zeta_3-{3\over 32}+{\pi^2\over 8}\right){1\over
N^2}\nonumber \\
&&
+\left({\pi^2\over 48}-{25\over 216}\right){(N^2-1)N_F\over N}\;,
\end{eqnarray}
so that
\begin{eqnarray}
\label{eq:Htwo}
H^{(2)} &=&  
\left(4\zeta_3+\frac{589}{432}- \frac{11\pi^2}{72}\right)N^2
+\left(-\frac{1}{2}\zeta_3-\frac{41}{54}-\frac{\pi^2}{48} \right)
+\left(-3\zeta_3 -\frac{3}{16} + \frac{\pi^2}{4}\right) \frac{1}{N^2}\nonumber \\
&&
+\left(-\frac{19}{18}+\frac{\pi^2}{36} \right) N\NF 
+\left(-\frac{1}{54}-\frac{\pi^2}{24}\right) \frac{\NF}{N}+ \frac{5}{27} \NF^2.
\end{eqnarray}
The factors $H^{(2)}_q$ and $H^{(2)}_g$ are directly related to those found 
in gluon--gluon scattering~\cite{m4}, quark--quark scattering~\cite{m2}
and quark--gluon scattering~\cite{m3} (which each involve four partons) 
as well as in the quark form factor~\cite{kl1,qff1,qff2,qff3}
and gluon form factor~\cite{harlander}.
We also note that (on purely dimensional grounds) one 
might expect terms of the 
type ${\tt S}_{ij}^2$ to be present in $H^{(2)}$.  Of course such terms are $1 + {\cal
O}(\epsilon)$ and therefore leave the pole part unchanged, only modifying
the finite remainder.
At present it is not known how to systematically include these effects.

\section{Helicity amplitudes}
\label{sec:helicity}
\setcounter{equation}{0}

We can extend the results of the previous section to include $Z$ boson exchange,
\begin{equation}
e^+(p_5) + e^-(p_6) \to \left(Z^*,\gamma^*\right)  (p_4) \longrightarrow q(p_1) + \bar q (p_2) + g(p_3)\; ,
\end{equation}
where the off-shell vector boson now distinguishes between left- and
right-handed fermions by keeping track of the helicity of the final state
quarks.\footnote{Note that the full matrix element for any process should be summed over 
both photon and $Z$-boson exchange.}
A convenient method to evaluate the helicity amplitudes is in terms
of  Weyl--van der Waerden spinors, which is described briefly in
Appendix~\ref{sec:appa} and in detail in \cite{WvdW,six}.

It is also straightforward to include the spin-correlations with the initial
state by contracting the hadronic current with the lepton current $V_\mu$ 
for fixed helicities of the initial state electron (and positron).
Using the spinor calculus of Appendix~\ref{sec:appa}
we can express the lepton current $V_\mu$ in terms 
of the helicities of the incident $e^+$ and $e^-$ (with momenta 
$p_5$ and $p_6$ respectively).
Explicitly, 
\begin{eqnarray}
V_\mu^\gamma(e^++,e^--) = e\sigma_\mu^{\dot AB}p_{6\dot 
A}p_{5B}\frac{L^\gamma_{ee}}{s},
& ~~~~~ &
V_\mu^Z(e^++,e^--) = e\sigma_\mu^{\dot AB}p_{6\dot 
A}p_{5B}\frac{L^Z_{ee}}{s-M_Z^2+i\Gamma_ZM_Z},\nonumber \\
V_\mu^\gamma(e^+-,e^-+) = e\sigma_\mu^{\dot AB}p_{5\dot 
A}p_{6B}\frac{R^\gamma_{ee}}{s},
& ~~~~~ &
V_\mu^Z(e^+-,e^-+) = e\sigma_\mu^{\dot AB}p_{5\dot 
A}p_{6B}\frac{R^Z_{ee}}{s-M_Z^2+i\Gamma_ZM_Z}.\nonumber \\
\end{eqnarray}

The hadronic current $\S_{\mu}$ is related to the fixed helicity 
currents, $\S_{{\dot A}B}$, by
\begin{equation}
\S_{\mu}(q+;g\lambda;\overline{q}-) = R^V_{f_1f_2}
 \sqrt2\, \sigma_{\mu}^{{\dot A}B} \S_{{\dot A}B}(q+;g\lambda;\overline{q}-) ,
\end{equation}
\begin{equation}
\S_{\mu}(q-;g\lambda;\overline{q}+) = L^V_{f_1f_2}
 \sqrt2\, \sigma_{\mu}^{{\dot A}B} \S_{{\dot A}B}(q-;g\lambda;\overline{q}+) .
\end{equation}
As in Eq.~(\ref{eq:renorme}), the gauge boson coupling is extracted 
from $\S_{{\dot A}B}$.
As mentioned earlier, the left- and right-handed currents couple with a
different strength when the vector boson is a $Z$.

The currents with the quark helicities flipped
follow from parity conservation:
\begin{equation}
\S_{{\dot A}B}(q-;g\lambda;\overline{q}+) =
( \S_{{\dot B}A}(q+;g(-\lambda);\overline{q}-))^*\ .
\end{equation}
Charge conjugation implies the following relations between currents with
different helicities:
\begin{equation}
\S_{{\dot 
A}B}(q\lambda_{q};g\lambda;\overline{q}\lambda_{\overline{q}}) =
(-1)
\S_{{\dot A}B}(\overline{q}\lambda_{\overline{q}};g\lambda;q\lambda_{q}).
\end{equation}
All helicity amplitudes are therefore related to the amplitudes with 
$\lambda_{q} = +$ and $\lambda_{\bar q} = -$.

Explicitly, we find
\begin{eqnarray}
{\S}_{\dot AB}(q+;g+;\overline{q}-)
&=&
\alpha(y,z)~
\frac{p_{1\dot AD}p_2^D p_{2B}}{\langle p_1p_3 \rangle\langle p_3{p}_2 \rangle}
+\beta(y,z)~
\frac{p_{3\dot AD} p_2^D p_{2B}}{\langle p_1p_3 \rangle\langle p_3{p}_2 \rangle}
+\gamma(y,z)~
\frac{p_{1\dot CB}p_3^{\dot C}p_{3\dot A}}{\langle p_1p_3 \rangle\langle p_3{p}_2 \rangle^*} \nonumber \\
&& +~\delta(y,z)~\frac{\langle p_1p_3\rangle^*}{\langle p_1p_3 \rangle\langle p_1{p}_2 \rangle^*}
\left(p_{1\dot AB}+p_{2\dot AB}+p_{3\dot AB}\right)
\;. 
\label{eq:helamp}
\end{eqnarray}
The other helicity amplitudes are obtained from $\S_{\dot 
AB}(q+;g+;\bar q-)$ by the above parity and charge conjugation relations,
while the coefficients $\alpha$, $\beta$ and $\gamma$ are written 
in terms of the tensor coefficients:
\begin{eqnarray}
\alpha(y,z) &=& \frac{\sbc\sac}{4}\bigg(2 B(\sac,\sbc,\sabc) + A_{12}(\sac,\sbc,\sabc) - A_{11}(\sac,\sbc,\sabc)\bigg),\nonumber \\
\beta(y,z) &=& \frac{\sac}{4}\bigg(2\sbc B(\sac,\sbc,\sabc) + 2 (\sab+\sac)A_{11}(\sac,\sbc,\sabc) \nonumber \\&& \hspace{0.7cm}
+\sbc\big(A_{12}(\sac,\sbc,\sabc)+A_{13}(\sac,\sbc,\sabc)\big)\bigg),\nonumber \\
\gamma(y,z) &=& \frac{\sac\sbc}{4}\bigg(A_{11}(\sac,\sbc,\sabc)-A_{13}(\sac,\sbc,\sabc)\bigg) ,\nonumber \\
\delta(y,z) &=&  - \frac{\sab\sac}{4}A_{11}(\sac,\sbc,\sabc).
\end{eqnarray}
When the hadron tensor is contracted with $\epsilon_4^\mu$ or the lepton current
$V^\mu$, the final term of Eq.~(\ref{eq:helamp}) vanishes\footnote{And for this
reason was omitted in Ref.~\cite{gg}.}.   Furthermore,
current conservation implies the following relation between the four helicity
coefficients,
\begin{equation}
\alpha(y,z)-\beta(y,z)-\gamma(y,z)-\frac{2\sabc}{\sab}~\delta(y,z) = 0.
\end{equation}
This relation is fulfilled automatically once the tensor coefficients are 
inserted and does therefore not yield a further reduction of the tensor basis.

As with the tensor coefficients, the helicity amplitude coefficients $\alpha$,
$\beta$ and $\gamma$ are vectors in colour space and
have perturbative expansions:
\begin{equation}
\Omega =  \sqrt{4\pi\alpha} \sqrt{4\pi\alpha_s} \; \bom{T}^a_{ij}\, \left[
\Omega^{(0)}  
+ \left(\frac{\alpha_s}{2\pi}\right) \Omega^{(1)}  
+ \left(\frac{\alpha_s}{2\pi}\right)^2 \Omega^{(2)} 
+ {\cal O}(\alpha_s^3) \right] \;,\nonumber \\
\end{equation}
for $\Omega = \alpha,\beta,\gamma$. The dependence on $(y,z)$ is again 
implicit.

The ultraviolet and infrared properties of the helicity coefficients match 
those of the tensor coefficients, 
\begin{eqnarray}
\Omega^{(0)}  &=& \Omega^{(0),{\rm un}} ,
 \nonumber \\
\Omega^{(1)}  &=& 
S_\e^{-1} \Omega^{(1),{\rm un}} 
-\frac{\beta_0}{2\e} \Omega^{(0),{\rm un}}  ,  \nonumber \\
\Omega^{(2)} &=& 
S_\e^{-2} \Omega^{(2),{\rm un}}  
-\frac{3\beta_0}{2\e} S_\e^{-1}
\Omega^{(1),{\rm un}}  
-\left(\frac{\beta_1}{4\e}-\frac{3\beta_0^2}{8\e^2}\right)
\Omega^{(0),{\rm un}},
\end{eqnarray}
and
\begin{eqnarray}
\Omega^{(1)} &=& {\bom I}^{(1)}(\epsilon) \Omega^{(0)} +
\Omega^{(1),{\rm finite}},\nonumber \\
\Omega^{(2)} &=& \Biggl (-\frac{1}{2}  {\bom I}^{(1)}(\epsilon) {\bom I}^{(1)}(\epsilon)
-\frac{\beta_0}{\epsilon} {\bom I}^{(1)}(\epsilon) 
+e^{-\epsilon \gamma } \frac{ \Gamma(1-2\epsilon)}{\Gamma(1-\epsilon)} 
\left(\frac{\beta_0}{\epsilon} + K\right)
{\bom I}^{(1)}(2\epsilon) + {\bom H}^{(2)}(\epsilon) 
\Biggr )\Omega^{(0)}\nonumber \\
&& + {\bom I}^{(1)}(\epsilon) \Omega^{(1)}+ \Omega^{(2),{\rm finite}},
\end{eqnarray}
where ${\bom I}^{(1)}(\epsilon)$ and ${\bom H}^{(2)}(\epsilon)$ are defined 
in Eqs.~(\ref{eq:I1}) and (\ref{eq:htwo}) respectively.

At leading order 
\begin{equation}
\alpha^{(0)}(y,z) = \beta^{(0)}(y,z) = 1\qquad
\mbox{and}\qquad \gamma^{(0)}(y,z) =0.
\end{equation}
The renormalized next-to-leading order helicity amplitude coefficients can be 
straightforwardly obtained to all orders in $\epsilon$ from 
the  tensor coefficients using Eqs.~(\ref{eq:aNLO})--(\ref{eq:bNLO}).
For practical purposes, they are needed through to ${\cal O}(\epsilon^2)$
in evaluating the 
infrared-divergent one-loop contribution to the two-loop amplitude,
while only the 
finite piece is needed for the one-loop self-interference.
They can be decomposed 
according to their colour structure as follows:
\begin{equation}
\Omega^{(1),{\rm finite}}(y,z) =  
N\, a_{\Omega}(y,z) + \frac{1}{N}\, b_{\Omega}(y,z) + \beta_0\, c_{\Omega}(y,z)
   \;. 
\label{eq:oneloopamp}
\end{equation}
The expansion of the coefficients 
through to $\e^2$ yields HPLs and 2dHPLs up to weight 4
for  $a_{\Omega}$, $b_{\Omega}$ and up to weight 3 for $c_{\Omega}$. The 
explicit expressions are of considerable size, such that we only quote the 
$\e^0$-terms here
(although these have been known already for a long time~\cite{gg}). 
The expressions through to ${\cal O}(\e^2)$ can be obtained in FORM format
from the authors. An example of the size and structure of those coefficients
can be found in~\cite{3jme}, where we explicitly list the helicity-averaged
one-loop times one-loop and tree times two-loop matrix elements. 
The one-loop coefficients read:
\begin{eqnarray}
a_{\alpha}(y,z)  &=& 
          - \frac{7}{4}
          - \frac{\pi^2}{12}
          + \frac{3}{8}\H(0;z)
          - \frac{1}{2} \H(0,z) \G(0;y)
          - \frac{1}{2} \H(1,0;z)
          - \frac{3}{8} \G(0;y)
          + \frac{1}{2} \G(1,0;y) \nonumber \\
&&       - \frac{1}{4(1-z)^2} \H(0;z) 
       - \frac{1}{4(1-z)} \Big(1+2\H(0;z)\Big) + {\cal O} (\e) \; ,
\nonumber \\
b_{\alpha}(y,z)  &=& 
         \frac{z^2}{2y^2} \Big(
             \H(0;z) \G(1 - z;y)
          +  \H(1;z) \G( - z;y)
          -  \G( - z,1 - z;y)
         \Big)
       + \frac{z}{2y}  \Big(
          - \H(0;z)\nonumber \\
&&
          + 2 \H(0;z) \G(1 - z;y)
          - \H(1;z)
          + 2 \H(1;z) \G( - z;y)
          + \G(1 - z;y)
          - 2 \G( - z,1 - z;y)
         \Big)\nonumber \\
&&
       + \frac{1}{2y (1-z)}\H(0;z)
       -\frac{1}{2y}  \H(0;z)
       + \frac{1}{4(1-z)^2}\H(0;z)
       + \frac{1}{4(1-z)}  \Big(
            1
          + 2 \H(0;z)
         \Big)
          + \frac{7}{4}
          - \frac{3}{4} \H(0;z)\nonumber \\
&&
          + \frac{1}{2} \H(0;z) \G(1 - z;y)
          + \frac{1}{2} \H(0,1;z)
          - \frac{3}{4} \H(1;z)
          + \H(1;z) \G( - z;y)
          - \frac{1}{2} \H(1;z) \G(0;y)\nonumber \\
&&
          + \frac{3}{4} \G(1 - z;y)
          + \frac{1}{2} \G(1 - z,0;y)
          - \G( - z,1 - z;y)
          + \frac{1}{2} \G(0,1 - z;y)
          - \frac{1}{2} \G(1,0;y)+ {\cal O} (\e)\;, 
\nonumber \\
c_{\alpha}(y,z)  &=& -\frac{1}{4} \H(0;z) - \frac{1}{4} \G(0;y) 
+ \frac{i\pi}{2}  + {\cal O} (\e)\;,  \nonumber \\
a_{\beta}(y,z)  &=& 
          - \frac{3}{2}
          - \frac{\pi^2}{12}
          + \frac{3}{8} \H(0;z)
          - \frac{1}{2} \H(0;z) \G(0;y)
          - \frac{1}{2} \H(1,0;z)
          - \frac{3}{8} \G(0;y)
          + \frac{1}{2} \G(1,0;y)\nonumber \\
&&
          +\frac{1}{4(1-z)} \H(0;z)+ {\cal O} (\e)\; ,
\nonumber \\
b_{\beta}(y,z)  &=& 
          \frac{z(1-z)}{2y^2} \Big(
          - \H(0;z) \G(1 - z;y)
          - \H(1;z) \G( - z;y)
          + \G( - z,1 - z;y)
          \Big)
       +  \frac{z}{2y}  \Big(
          -  \H(0;z)\nonumber \\
&&
          + 2\H(0;z) \G(1 - z;y)
          - \H(1;z)
          + 2\H(1;z) \G( - z;y)
          + \G(1 - z;y)
          - 2\G( - z,1 - z;y)
          \Big)\nonumber \\
&&
       +  \frac{1}{2y}  \Big(
          - 2\H(0;z) \G(1 - z;y)
          +  \H(1;z)
          - 2\H(1;z) \G( - z;y)
          -  \G(1 - z;y)
          + 2\G( - z,1 - z;y)
          \Big)\nonumber \\
&&
       +  \frac{z}{2(y+z)^2}  \Big(
          -  \H(1;z)
          +  \G(1 - z;y)
          \Big)
       +  \frac{z}{2(y+z)}
       -  \frac{1}{4(1-z)}  \H(0;z)
       +  \frac{1}{2(y+z)}  \Big(
              \H(1;z)\nonumber \\
&&
          -   \G(1 - z;y)
          \Big)
          + \frac{3}{2}
          - \frac{3}{4} \H(0;z)
          + \frac{1}{2} \H(0;z) \G(1 - z;y)
          + \frac{1}{2} \H(0,1;z)
          - \frac{3}{4} \H(1;z)\nonumber \\
&&
          + \H(1;z) \G( - z;y)
          - \frac{1}{2} \H(1;z) \G(0;y)
          + \frac{3}{4} \G(1 - z;y)
          + \frac{1}{2} \G(1 - z,0;y)
          - \G( - z,1 - z;y)\nonumber \\
&&
          + \frac{1}{2} \G(0,1 - z;y)
          - \frac{1}{2} \G(1,0;y)
          \Big)+ {\cal O} (\e) \;,
\nonumber \\
c_{\beta}(y,z)  &=& -\frac{1}{4} \H(0;z) - \frac{1}{4} \G(0;y) 
+ \frac{i\pi}{2} + {\cal O} (\e) \;,\nonumber \\
a_{\gamma}(y,z)  &=& -\frac{1}{4} + \frac{1}{4(1-z)^2} \H(0;z) 
              + \frac{1}{4(1-z)} \Big( 1-\H(0;z)\Big)  + {\cal O} (\e)
\;,\nonumber \\
b_{\gamma}(y,z)  &=&
\frac{1}{4} + \frac{z}{2y^2}\Big( 
          - \H(0;z) \G(1 - z;y)
          - \H(1;z) \G( - z;y)
          + \G( - z,1 - z;y)
          \Big)
       - \frac{1}{2y(1-z)} \H(0;z)\nonumber \\
&&
       + \frac{1}{2y} \Big(
            \H(0;z)
          + \H(1;z)
          - \G(1 - z;y)
          \Big)
       + \frac{z}{2(y+z)^2} \Big(
            \H(1;z)
          - \G(1 - z;y)
          \Big)
       - \frac{z}{2(y+z)}\nonumber \\
&&
       + \frac{1}{2(y+z)} \Big(
          - \H(1;z)
          + \G(1 - z;y)
          \Big)
       - \frac{1}{4(1-z)^2}\H(0;z)
       + \frac{1}{4(1-z)}\Big(
          - 1
          + \H(0;z)
          \Big) + {\cal O} (\e)\; , 
 \nonumber \\
c_{\gamma}(y,z)  &=& 0 \; . 
\label{eq:omega1}
\end{eqnarray}
It should be noted that these finite pieces of the one-loop coefficients 
can equally well be written in terms of ordinary logarithms and dilogarithms,
see~\cite{ert1,gg}. The reason to express them in terms of HPLs and 2dHPLs 
here is their usage in the infrared counter-term of the two-loop coefficients, 
which cannot be fully expressed in terms of logarithmic and polylogarithmic 
functions. 

The finite two-loop remainder is obtained by subtracting the
predicted infrared structure (expanded through to ${\cal O}(\epsilon^0)$) from
the renormalized helicity coefficient.  We further decompose the 
finite remainder according to the colour structure, as follows:
\begin{eqnarray}
\Omega^{(2),{\rm finite}}(y,z) &=&  
N^2 A_\Omega(y,z) + B_\Omega(y,z) + \frac{1}{N^2} C_\Omega(y,z) 
+ N\NF D_\Omega(y,z) \nonumber \\ &&
+ \frac{\NF}{N} E_\Omega(y,z) + \NF^2 F_\Omega (y,z)
+ \NFZ \left(\frac{4}{N}-N\right) G_\Omega(y,z) \; , 
\label{eq:twoloopamp}
\end{eqnarray}
where the last term is generated by graphs where the virtual gauge boson does not
couple directly to the final-state quarks.   This contribution is denoted
by $\NFZ$ and is proportional
to the charge weighted  sum of the quark flavours. 
In the case of purely electromagnetic interactions we find,
\begin{equation}
N_{F,\gamma} = \frac{\sum_q e_q}{e_q}\; .
\end{equation}
Including $Z$-interactions, the same class of diagrams yields not only a 
contribution from the  vector component of the $Z$, which for the right-handed
quark amplitude is given by
\begin{equation}
N_{F,Z} = \frac{\sum_q \left(L^Z_{qq}+R^Z_{qq}\right)}{2R^Z_{qq}}\; ,
\end{equation}
 but 
also a contribution involving the axial couplings of the $Z$~\cite{kk}. This 
contribution vanishes if summed over isospin doublets. The 
large mass splitting of the third quark family induces a non-vanishing 
contribution from this class of diagrams, which can however not be computed 
within the framework of massless QCD employed here, but can only be 
obtained within an effective theory with large top-quark mass. 
In contrast to the vector contribution from these diagrams, which is 
finite, one could expect divergences in the axial vector contribution, which 
would be cancelled by the single unresolved limits of the 
corresponding axial contributions to four-parton final 
states~\cite{onel4p1,onel4p2}. Results from the four-parton final states 
show that this axial contribution is numerically very 
small~\cite{menloparc}. 

The helicity coefficients contain HPLs and 2dHPLs up to weight 4 
in the $A,B,C,G$-terms, up to weight 3 in the $D,E$-terms 
(which do moreover contain only a limited subset of purely planar master 
integrals) and up to 
weight 2 in the $F$-term. The size of each helicity coefficient is comparable 
to the size of the 
helicity-averaged tree times two-loop matrix element quoted in~\cite{3jme}.
We do therefore only quote the $A$- and $D$-terms of each coefficient, which 
form the leading colour contributions, and which turn out to be numerically 
dominant, approximating the full expressions to an accuracy of about 20\%. 
The complete set of coefficients in FORM format can be obtained from the 
authors. 

These leading colour terms are:
\begin{eqnarray}
\lefteqn{A_{\alpha}(y,z) = }\nonumber \\
&&          \frac{1}{48y(1-z)}  \Big[
          \pi^2
          - 13 \H(0;z)
          + 6 \H(1,0;z)
          + 6 \G(1,0;y)
         \Big]
       -  \frac{1}{48y(1-y-z)}  \Big[
          \pi^2
          - 13 \H(0;z)
          + 6 \H(1,0;z)
\nonumber \\ &&
          + 6 \G(1,0;y)
         \Big]
       -  \frac{z}{16(1-y)^2} \G(0;y)
       -  \frac{z}{16(1-y)}
       +  \frac{z}{12(1-y-z)^2} \Big[
          - \frac{5\pi^2}{6}
          - 5 \H(0;z) \G(0;y)
          - 5 \H(1,0;z)
\nonumber \\ &&
          + 5 \G(1,0;y)
         \Big]
       +  \frac{z}{16(1-y-z)} \Big[
            \frac{14\pi^2}{3}
          - 11 \H(0;z)
          + 28 \H(0;z) \G(0;y)
          + 28 \H(1,0;z)
          + 11 \G(0;y)
\nonumber \\ &&
          - 28 \G(1,0;y)
         \Big]
       +  \frac{z^2}{16(1-y-z)^2}  \Big[
            \frac{11\pi^2}{6}
          + 11 \H(0;z) \G(0;y)
          + 11 \H(1,0;z)
          - 11 \G(1,0;y)
         \Big]
\nonumber \\ &&
       +  \frac{1}{3(1-y)} \G(0;y)
       +  \frac{1}{48(1-z)^2}  \Big[
          - \frac{\pi^2}{6}
          + \pi^2 \big(3 \H(0;z)
                     +3 \H(1;z)
                     - \G(1 - z;y)
                     + \G(0;y)\big)
          + 6\zeta_3
\nonumber \\ &&
          - \frac{355}{6} \H(0;z)
          - 6\H(0;z) \G(1 - z,0;y)
          + 10 \H(0;z) \G(0;y)
          + 45 \H(0,0;z)
          + 12 \H(0,0;z) \G(0;y)
\nonumber \\ &&
          + 18 \H(0,1,0;z)
          -  \H(1,0;z)
          - 6 \H(1,0;z) \G(1 - z;y)
          + 6 \H(1,0;z) \G(0;y)
          + 12 \H(1,0,0;z)
          + 18 \H(1,1,0;z)
\nonumber \\ &&
          + 6 \G(1 - z,1,0;y)
          - 6 \G(0,1,0;y)
         \Big]
       + \frac{1}{72(1-z)}\Big[
          \pi^2 \big(  
          - 8
          + 9 \H(0;z)
          + 9 \H(1;z)
          - 3 \G(1 - z;y)
\nonumber \\ &&
          + 3 \G(0;y)
                \big)
          + 18\zeta_3
          - \frac{277}{4}
          - 65 \H(0;z)
          - 18 \H(0;z) \G(1 - z,0;y)
          + 39 \H(0;z) \G(0;y)
          + 81 \H(0,0;z)
\nonumber \\ &&
          + 36 \H(0,0;z) \G(0;y)
          + 54 \H(0,1,0;z)
          - 48 \H(1,0;z)
          - 18 \H(1,0;z) \G(1 - z;y)
          + 18 \H(1,0;z) \G(0;y)
\nonumber \\ &&
          + 36 \H(1,0,0;z)
          + 54 \H(1,1,0;z)
          + 18 \G(1 - z,1,0;y)
          + 15 \G(0;y)
          - 18 \G(0,1,0;y)
          - 9 \G(1,0;y)
         \Big]
\nonumber \\ &&
       +  \frac{1}{48(1-y-z)^2} \Big[
          - 2\pi^2 \H(1;z)
          - \pi^2 \G(0;y)
          + 12 \zeta_3
          - 6 \H(1,0;z) \G(0;y)
          - 12 \H(1,1,0;z)
          - 6 \G(0,1,0;y)
         \Big]
\nonumber \\ &&
       +  \frac{1}{48(1-y-z)} \Big[
          - 4\pi^2 \H(1;z)
          - 2\pi^2 \G(0;y)
          + 24 \zeta_3  
          - 13 \H(0;z)
          - 12 \H(1,0;z) \G(0;y)
          - 24 \H(1,1,0;z)
\nonumber \\ &&
          - 20 \G(0;y)
          - 12 \G(0,1,0;y)
         \Big]
       +  \frac{\pi^2}{288}  \Big[
          - \frac{928}{3}
          - 5 \H(0;z)
          + 12 \H(0;z) \G(1 - z;y)
          + 36 \H(0;z) \G(0;y)
\nonumber \\ &&
          - 12 \H(0;z) \G(1;y)
          + 24 \H(0,1;z)
          + 24 \H(1;z) \G(1 - z;y)
          - 24 \H(1;z) \G( - z;y)
          - 12 \H(1;z) \G(1;y)
\nonumber \\ &&
          + 24 \H(1,0;z)
          + 12 \H(1,1;z)
          - 44 \G(1 - z;y)
          + 12 \G(1 - z,0;y)
          - 24 \G(1 - z,1;y)
          + 24 \G( - z,1 - z;y)
\nonumber \\ &&
          - 24 \G(0,1 - z;y)
          + 49 \G(0;y)
          - 24 \G(0,1;y)
          + 12 \G(1,1 - z;y)
          + 44 \G(1;y)
          - 36 \G(1,0;y)
          + 24 \G(1,1;y)
         \Big]   
\nonumber \\ &&
       +  \frac{\zeta_3}{72}  \Big[
            317
          - 18 \H(0;z)
          + 90 \H(1;z)
          - 72 \G(1 - z;y)
          - 18 \G(0;y)
          - 18 \G(1;y)
         \Big]
       +  \frac{11\pi^4}{360}
       +  \frac{1}{72} \Big[
          - \frac{89959}{144}
\nonumber \\ &&
          + \frac{2149}{12} \H(0;z)
          - 66 \H(0;z) \G(1 - z,0;y)
          - 18 \H(0;z) \G(1 - z,1,0;y)
          + 36 \H(0;z) \G( - z,1 - z,0;y)
\nonumber \\ &&
          - 36 \H(0;z) \G(0,1 - z,0;y)
          - 66 \H(0;z) \G(0;y)
          + 126 \H(0;z) \G(0,0;y)
          - 18 \H(0;z) \G(0,1,0;y)
\nonumber \\ &&
          + 18 \H(0;z) \G(1,1 - z,0;y)
          - 3 \H(0;z) \G(1,0;y)
          - 36 \H(0;z) \G(1,0,0;y)
          + \frac{23}{2} \H(0,0;z)
\nonumber \\ &&
          + 72 \H(0,0;z) \G(0;y)
          + 36 \H(0,0;z) \G(0,0;y)
          + 72 \H(0,0,1,0;z)
          + 3 \H(0,1,0;z)
          - 18 \H(0,1,0;z) \G(1 - z;y)
\nonumber \\ &&
          + 36 \H(0,1,0;z) \G( - z;y)
          + 18 \H(0,1,0;z) \G(0;y)
          - 18 \H(0,1,0;z) \G(1;y)
          + 36 \H(0,1,1,0;z)
\nonumber \\ &&
          - 71 \H(1,0;z)
          - 66 \H(1,0;z) \G(1 - z;y)
          + 18 \H(1,0;z) \G(1 - z,0;y)
          + 36 \H(1,0;z) \G( - z,1 - z;y)
\nonumber \\ &&
          - 36 \H(1,0;z) \G( - z,0;y)
          - 36 \H(1,0;z) \G(0,1 - z;y)
          + 96 \H(1,0;z) \G(0;y)
          + 18 \H(1,0;z) \G(1,1 - z;y)
\nonumber \\ &&
          - 18 \H(1,0;z) \G(1,0;y)
          + 72 \H(1,0,0;z)
          + 36 \H(1,0,0;z) \G(0;y)
          + 72 \H(1,0,1,0;z)
\nonumber \\ &&
          + 36 \H(1,1,0;z) \G(1 - z;y)
          - 36 \H(1,1,0;z) \G( - z;y)
          - 18 \H(1,1,0;z) \G(1;y)
          + 36 \H(1,1,0,0;z)
\nonumber \\ &&
          + 18 \H(1,1,1,0;z)
          + 18 \G(1 - z,0,1,0;y)
          + 66 \G(1 - z,1,0;y)
          + 36 \G(1 - z,1,1,0;y)
\nonumber \\ &&
          - 36 \G( - z,1 - z,1,0;y)
          - 36 \G( - z,0,1,0;y)
          + 36 \G(0,1 - z,1,0;y)
          + \frac{49}{3} \G(0;y)
          + 160 \G(0,0;y)
\nonumber \\ &&
          - 36 \G(0,0,1,0;y)
          - 30 \G(0,1,0;y)
          + 36 \G(0,1,1,0;y)
          - 18 \G(1,1 - z,1,0;y)
          + 71 \G(1,0;y)
\nonumber \\ &&
          - 126 \G(1,0,0;y)
          + 54 \G(1,0,1,0;y)
          - 66 \G(1,1,0;y)
          + 36 \G(1,1,0,0;y)
          - 36 \G(1,1,1,0;y)
         \Big]
\nonumber \\ &&
 + i\pi \Bigg\{ 
          - \frac{11}{16(1-z)^2} \H(0;z)
       + \frac{1}{16(1-z)} \Big[
          - 11
          - 22 \H(0;z)
          \Big]
       + 2\zeta_3
       + \frac{1}{48} \Big[
          - \frac{44\pi^2}{3}
          - \frac{2345}{18}
          - 11 \H(0;z)
\nonumber \\ && \hspace{9mm}
          - 66 \H(0;z) \G(0;y)
          - 66 \H(1,0;z)
          - 110 \G(0;y)
          + 66 \G(1,0;y)
           \Big] \Bigg\}
\nonumber \\
\lefteqn{D_{\alpha}(y,z) =} \nonumber \\&&       
       \frac{1}{12y(1-z)}\H(0;z)  
       - \frac{1}{12y(1-y-z)}\H(0;z)
       + \frac{z}{6(1-y-z)^2} \Big[ 
           \frac{\pi^2}{6} 
          + \H(0;z) \G(0;y)
          + \H(1,0;z)
\nonumber \\ &&
          - \G(1,0;y)
         \Big]
       +  \frac{z}{4(1-y-z)} \Big[
          - \frac{\pi^2}{3}
          + \H(0;z)
          - 2\H(0;z) \G(0;y)
          - 2\H(1,0;z)
          - \G(0;y)
          + 2\G(1,0;y)
         \Big]
\nonumber \\ &&
       +  \frac{z^2}{4(1-y-z)^2}  \Big[
          - \frac{\pi^2}{6}
          - \H(0;z) \G(0;y)
          - \H(1,0;z)
          + \G(1,0;y)
         \Big]
       - \frac{1}{12(1-y)}\G(0;y)
\nonumber \\&&
       +  \frac{1}{72(1-z)^2} \Big[
            \pi^2
          + 25 \H(0;z)
          - \frac{3}{2} \H(0;z) \G(0;y)
          - 9 \H(0,0;z)
          + 6 \H(1,0;z)
         \Big]
\nonumber \\&&
       +  \frac{1}{144(1-z)}  \Big[
            4\pi^2
          + 38
          + 37 \H(0;z)
          - 6 \H(0;z) \G(0;y)
          - 36 \H(0,0;z)
          + 24 \H(1,0;z)
          - 3 \G(0;y)
         \Big]
\nonumber \\&&
       +  \frac{1}{12(1-y-z)}  \Big[
            \H(0;z)
          + 2 \G(0;y)
         \Big]
       +  \frac{\pi^2}{72}  \Big[
            \frac{395}{12}
          -  \H(0;z)
          + 2 \G(1 - z;y)
          -  \G(0;y)
          - 2 \G(1;y)
         \Big]
\nonumber \\&&
       - \frac{19}{36} \zeta_3 
       +  \frac{1}{144}  \Big[
            \frac{3661}{18}
          - 25 \H(0;z)
          + 24 \H(0;z) \G(1 - z,0;y)
          + 29 \H(0;z) \G(0;y)
          - 36 \H(0;z) \G(0,0;y)
\nonumber \\&&
          + 6 \H(0;z) \G(1,0;y)
          - 28 \H(0,0;z)
          - 36 \H(0,0;z) \G(0;y)
          - 6 \H(0,1,0;z)
          + 40 \H(1,0;z)
\nonumber \\&&
          + 24 \H(1,0;z) \G(1 - z;y)
          - 30 \H(1,0;z) \G(0;y)
          - 36 \H(1,0,0;z)
          - 24 \G(1 - z,1,0;y)
          + 53 \G(0;y)
\nonumber \\&&
          - 82 \G(0,0;y)
          + 6 \G(0,1,0;y)
          - 40 \G(1,0;y)
          + 36 \G(1,0,0;y)
          + 24 \G(1,1,0;y)
         \Big]
\nonumber \\&&
 + i\pi \Bigg\{ 
         \frac{1}{8(1-z)^2} \H(0;z)
       + \frac{1}{8(1-z)} \Big[
            1
          + 2 \H(0;z)
          \Big]
       + \frac{1}{48} \Big[
            \frac{8\pi^2}{3} 
          - \frac{28}{3}
          + 13 \H(0;z)
          + 12 \H(0;z) \G(0;y)
\nonumber \\&& \hspace{9mm}
          + 12 \H(1,0;z)
          + 31 \G(0;y)
          - 12 \G(1,0;y)
          \Big] \Bigg\}
    \nonumber \\ 
\lefteqn{A_{\beta}(y,z) =} \nonumber \\ &&
         - \frac{z}{16(1-y)^2} \G(0;y)
         - \frac{z}{16(1-y)}
      + \frac{z}{16(y+z)^2} \Big[
            \frac{47\pi^2}{3} \H(1;z)
          - \frac{47\pi^2}{3} \G(1 - z;y)
          - 94 \H(0;z) \G(1 - z,0;y)
\nonumber \\&&
          - 94 \H(0,1,0;z)
          - 99 \H(1,0;z)
          - 94 \H(1,0;z) \G(1 - z;y)
          + 94 \H(1,0;z) \G(0;y)
          + 94 \H(1,1,0;z)
\nonumber \\&&
          + 94 \G(1 - z,1,0;y)
          + 94 \G(0,1,0;y)
          - 99 \G(1,0;y)
         \Big]
       +  \frac{z}{16(y+z)}  \Big[
          - \frac{47\pi^2}{3}
          + 11
          + 44 \H(0;z)
\nonumber \\&&
          - 94 \H(0;z) \G(0;y)
          - 94 \H(1,0;z)
          - 55 \G(0;y)
          + 94 \G(1,0;y)
         \Big]
       +  \frac{z}{12(1-y-z)^2} \Big[
            \frac{5\pi^2}{6} 
          - \frac{\pi^2}{2} \H(1;z) 
\nonumber \\&&
          - \frac{\pi^2}{4} \G(0;y)
          + 3\zeta_3
          + 5 \H(0;z) \G(0;y)
          + 5 \H(1,0;z)
          - \frac{3}{2} \H(1,0;z) \G(0;y)
          - 3 \H(1,1,0;z)
          - \frac{3}{2} \G(0,1,0;y)
\nonumber \\ &&
          - 5 \G(1,0;y)
         \Big] 
       +  \frac{z}{12(1-y-z)}  \Big[
          - \frac{19\pi^2}{3}
          + 5 \H(0;z)
          - 38 \H(0;z) \G(0;y)
          - 38 \H(1,0;z)
          - \frac{53}{4} \G(0;y)
\nonumber \\ &&
          + 38 \G(1,0;y)
         \Big]
       +  \frac{z^2}{8(y+z)^3}  \Big[
          - 11\pi^2 \H(1;z)
          + 11\pi^2 \G(1 - z;y)
          + 66 \H(0;z) \G(1 - z,0;y)
          + 66 \H(0,1,0;z)
\nonumber \\ &&
          + 33 \H(1,0;z)
          + 66 \H(1,0;z) \G(1 - z;y)
          - 66 \H(1,0;z) \G(0;y)
          - 66 \H(1,1,0;z)
          - 66 \G(1 - z,1,0;y)
\nonumber \\ &&
          - 66 \G(0,1,0;y)
          + 33 \G(1,0;y)
         \Big]
       +  \frac{z^2}{16(y+z)^2}  \Big[
          + 22\pi^2
          - 33 \H(0;z)
          + 132 \H(0;z) \G(0;y)
          + 132 \H(1,0;z)
\nonumber \\ &&
          + 33 \G(0;y)
          - 132 \G(1,0;y)
         \Big]
       +  \frac{z^2}{16(y+z)}  \Big[
            11\pi^2
          - 11 \H(0;z)
          + 66 \H(0;z) \G(0;y)
          + 66 \H(1,0;z)
\nonumber \\ &&
          + 11 \G(0;y)
          - 66 \G(1,0;y)
         \Big]
       +  \frac{z^2}{48(1-y-z)^2}  \Big[
          - \frac{53\pi^2}{6}
          - 53 \H(0;z) \G(0;y)
          - 53 \H(1,0;z)
          + 53 \G(1,0;y)
         \Big]
\nonumber \\ &&
       +  \frac{z^2}{16(1-y-z)} \Big[
            11\pi^2
          - 11 \H(0;z)
          + 66 \H(0;z) \G(0;y)
          + 66 \H(1,0;z)
          + 11 \G(0;y)
          - 66 \G(1,0;y)
         \Big]
\nonumber \\ &&
       +  \frac{z^3}{8(y+z)^4} \Big[
            \frac{11\pi^2}{2} \H(1;z)
          - \frac{11\pi^2}{2} \G(1 - z;y)
          - 33 \H(0;z) \G(1 - z,0;y)
          - 33 \H(0,1,0;z)
\nonumber \\ &&
          - 33 \H(1,0;z) \G(1 - z;y)
          + 33 \H(1,0;z) \G(0;y)
          + 33 \H(1,1,0;z)
          + 33 \G(1 - z,1,0;y)
          + 33 \G(0,1,0;y)
         \Big]
\nonumber \\ &&
       +  \frac{z^3}{8(y+z)^3} \Big[
          - \frac{11\pi^2}{2}
          - 33 \H(0;z) \G(0;y)
          - 33 \H(1,0;z)
          + 33 \G(1,0;y)
         \Big]
       +  \frac{z^3}{16(y+z)^2} \Big[
          - \frac{11\pi^2}{2}
\nonumber \\ &&
          - 33 \H(0;z) \G(0;y)
          - 33 \H(1,0;z)
          + 33 \G(1,0;y)
         \Big]
       +  \frac{z^3}{8(y+z)}  \Big[
          - \frac{11\pi^2}{6}
          - 11 \H(0;z) \G(0;y)
          - 11 \H(1,0;z)
\nonumber \\ &&
          + 11 \G(1,0;y)
         \Big]
       +  \frac{z^3}{16(1-y-z)^2} \Big[
            \frac{11\pi^2}{6}
          + 11 \H(0;z) \G(0;y)
          + 11 \H(1,0;z)
          - 11 \G(1,0;y)
         \Big]
\nonumber \\ &&
       +  \frac{z^3}{8(1-y-z)}  \Big[
          - \frac{11\pi^2}{6}
          - 11 \H(0;z) \G(0;y)
          - 11 \H(1,0;z)
          + 11 \G(1,0;y)
         \Big]
       +  \frac{5}{24(1-y)} \G(0;y)
\nonumber \\ &&
       +  \frac{1}{48(1-z)} \Big[
           \pi^2 \big(
            \frac{1}{6}
          - 3 \H(0;z)
          - 3 \H(1;z)
          +  \G(1 - z;y)
          -  \G(0;y)
         \big)
          - 6\zeta_3
          + \frac{355}{6} \H(0;z)
 \nonumber \\ &&
          + 6 \H(0;z) \G(1 - z,0;y)
          - 10 \H(0;z) \G(0;y)
          - 45 \H(0,0;z)
          - 12 \H(0,0;z) \G(0;y)
          - 18 \H(0,1,0;z)
 \nonumber \\ &&
          + \H(1,0;z)
          + 6 \H(1,0;z) \G(1 - z;y)
          - 6 \H(1,0;z) \G(0;y)
          - 12 \H(1,0,0;z)
          - 18 \H(1,1,0;z)
 \nonumber \\ &&
          - 6 \G(1 - z,1,0;y)
          + 6 \G(0,1,0;y)
         \Big]
       +  \frac{1}{8(y+z)} \Big[
          - \frac{7\pi^2}{3} \H(1;z)
          + \frac{7\pi^2}{3} \G(1 - z;y)
          + 14 \H(0;z) \G(1 - z,0;y)
 \nonumber \\ &&
          + 14 \H(0,1,0;z)
          + 25 \H(1,0;z)
          + 14 \H(1,0;z) \G(1 - z;y)
          - 14 \H(1,0;z) \G(0;y)
          - 14 \H(1,1,0;z)
\nonumber \\ &&
          - 14 \G(1 - z,1,0;y)
          - 14 \G(0,1,0;y)
          + 25 \G(1,0;y)
         \Big] 
       +  \frac{1}{8(1-y-z)^2}  \Big[
            \frac{\pi^2}{3} \H(1;z)
          + \frac{\pi^2}{6} \G(0;y)
          - 2\zeta_3
 \nonumber \\ &&
          +   \H(1,0;z) \G(0;y)
          + 2 \H(1,1,0;z)
          +  \G(0,1,0;y)
         \Big]
       +  \frac{1}{24(1-y-z)} \Big[
            \frac{13\pi^2}{6}
          + 10 \H(0;z) \G(0;y)
          + 13 \H(1,0;z)
\nonumber \\ &&
          + 10 \G(0;y)
          - 7 \G(1,0;y)
         \Big]
       +  \frac{\pi^2}{288}  \Big[
          - \frac{946}{3}
          - 5 \H(0;z)
          + 12 \H(0;z) \G(1 - z;y)
          + 36 \H(0;z) \G(0;y)
\nonumber \\ &&
          - 12 \H(0;z) \G(1;y)
          + 24 \H(0,1;z)
          + 24 \H(1;z) \G(1 - z;y)
          - 24 \H(1;z) \G( - z;y)
          - 12 \H(1;z) \G(1;y)
\nonumber \\ &&
          + 24 \H(1,0;z)
          + 12 \H(1,1;z)
          - 44 \G(1 - z;y)
          + 12 \G(1 - z,0;y)
          - 24 \G(1 - z,1;y)
          + 24 \G( - z,1 - z;y)
\nonumber \\ &&
          - 24 \G(0,1 - z;y)
          + 49 \G(0;y)
          - 24 \G(0,1;y)
          + 12 \G(1,1 - z;y)
          + 44 \G(1;y)
          - 36 \G(1,0;y)
          + 24 \G(1,1;y)
         \Big]
\nonumber \\ &&
       +  \frac{11\pi^4}{360} 
       +  \frac{\zeta_3}{72}  \Big[
            317
          - 18 \H(0;z)
          + 90 \H(1;z)
          - 72 \G(1 - z;y)
          - 18 \G(0;y)
          - 18 \G(1;y)
         \Big]
       +  \frac{1}{144} \Big[
          - \frac{79987}{72}
\nonumber \\ &&
          + \frac{1735}{6} \H(0;z)
          - 132 \H(0;z) \G(1 - z,0;y)
          - 36 \H(0;z) \G(1 - z,1,0;y)
          + 72 \H(0;z) \G( - z,1 - z,0;y)
\nonumber \\ &&
          - 72 \H(0;z) \G(0,1 - z,0;y)
          - 150 \H(0;z) \G(0;y)
          + 252 \H(0;z) \G(0,0;y)
          - 36 \H(0;z) \G(0,1,0;y)
\nonumber \\ &&
          + 36 \H(0;z) \G(1,1 - z,0;y)
          - 6 \H(0;z) \G(1,0;y)
          - 72 \H(0;z) \G(1,0,0;y)
          + 23 \H(0,0;z)
\nonumber \\ &&
          + 144 \H(0,0;z) \G(0;y)
          + 72 \H(0,0;z) \G(0,0;y)
          + 144 \H(0,0,1,0;z)
          + 6 \H(0,1,0;z)
\nonumber \\ &&
          - 36 \H(0,1,0;z) \G(1 - z;y)
          + 72 \H(0,1,0;z) \G( - z;y)
          + 36 \H(0,1,0;z) \G(0;y)
          - 36 \H(0,1,0;z) \G(1;y)
\nonumber \\ &&
          + 72 \H(0,1,1,0;z)
          - 160 \H(1,0;z)
          - 132 \H(1,0;z) \G(1 - z;y)
          + 36 \H(1,0;z) \G(1 - z,0;y)
\nonumber \\ &&
          + 72 \H(1,0;z) \G( - z,1 - z;y)
          - 72 \H(1,0;z) \G( - z,0;y)
          - 72 \H(1,0;z) \G(0,1 - z;y)
          + 192 \H(1,0;z) \G(0;y)
\nonumber \\ &&
          + 36 \H(1,0;z) \G(1,1 - z;y)
          - 36 \H(1,0;z) \G(1,0;y)
          + 144 \H(1,0,0;z)
          + 72 \H(1,0,0;z) \G(0;y)
\nonumber \\ &&
          + 144 \H(1,0,1,0;z)
          + 72 \H(1,1,0;z) \G(1 - z;y)
          - 72 \H(1,1,0;z) \G( - z;y)
          - 36 \H(1,1,0;z) \G(1;y)
\nonumber \\ &&
          + 72 \H(1,1,0,0;z)
          + 36 \H(1,1,1,0;z)
          + 36 \G(1 - z,0,1,0;y)
          + 132 \G(1 - z,1,0;y)
          + 72 \G(1 - z,1,1,0;y)
\nonumber \\ &&
          - 72 \G( - z,1 - z,1,0;y)
          - 72 \G( - z,0,1,0;y)
          + 72 \G(0,1 - z,1,0;y)
          + \frac{8}{3} \G(0;y)
          + 320 \G(0,0;y)
\nonumber \\ &&
          - 72 \G(0,0,1,0;y)
          - 60 \G(0,1,0;y)
          + 72 \G(0,1,1,0;y)
          - 36 \G(1,1 - z,1,0;y)
          + 160 \G(1,0;y)
\nonumber \\ &&
          - 252 \G(1,0,0;y)
          + 108 \G(1,0,1,0;y)
          - 132 \G(1,1,0;y)
          + 72 \G(1,1,0,0;y)
          - 72 \G(1,1,1,0;y)
         \Big] 
\nonumber \\ &&
 + i\pi \Bigg\{
         \frac{11}{16(1-z)} \H(0;z) 
       + 2\zeta_3
       + \frac{1}{48} \Big[
          - \frac{44\pi^2}{3}
          - \frac{1751}{18}
          - 11 \H(0;z)
          - 66 \H(0;z) \G(0;y)
          - 66 \H(1,0;z)
\nonumber \\ && \hspace{9mm}
          - 110 \G(0;y)
          + 66 \G(1,0;y)
          \Big]\Bigg\}
\nonumber \\
\lefteqn{D_{\beta}(y,z) =} \nonumber \\ &&
          \frac{z}{4(y+z)^2} \Big[
          - \frac{4\pi^2}{3} \H(1;z)
          + \frac{4\pi^2}{3} \G(1 - z;y)
          + 8 \H(0;z) \G(1 - z,0;y)
          + 8 \H(0,1,0;z)
          + 9 \H(1,0;z)
\nonumber \\ &&
          + 8 \H(1,0;z) \G(1 - z;y)
          - 8 \H(1,0;z) \G(0;y)
          - 8 \H(1,1,0;z)
          - 8 \G(1 - z,1,0;y)
          - 8 \G(0,1,0;y)
\nonumber \\ &&
          + 9 \G(1,0;y)
         \Big]
       +  \frac{z}{4(y+z)} \Big[
           \frac{4\pi^2}{3} 
          - 1
          - 4 \H(0;z)
          + 8 \H(0;z) \G(0;y)
          + 8 \H(1,0;z)
          + 5 \G(0;y)
          - 8 \G(1,0;y)
         \Big]
\nonumber \\ &&
       +  \frac{z}{6(1-y-z)^2}  \Big[
          - \frac{\pi^2}{6}
          -  \H(0;z) \G(0;y)
          -  \H(1,0;z)
          +  \G(1,0;y)
         \Big]
       +  \frac{z}{12(1-y-z)}  \Big[
            \frac{7\pi^2}{3}
          - 2 \H(0;z)
\nonumber \\ &&
          + 14 \H(0;z) \G(0;y)
          + 14 \H(1,0;z)
          + 5 \G(0;y)
          - 14 \G(1,0;y)
         \Big]
       +  \frac{z^2}{2(y+z)^3}  \Big[
            \pi^2 \H(1;z)
          - \pi^2 \G(1 - z;y)
\nonumber \\ &&
          - 6 \H(0;z) \G(1 - z,0;y)
          - 6 \H(0,1,0;z)
          - 3 \H(1,0;z)
          - 6 \H(1,0;z) \G(1 - z;y)
          + 6 \H(1,0;z) \G(0;y)
\nonumber \\ &&
          + 6 \H(1,1,0;z)
          + 6 \G(1 - z,1,0;y)
          + 6 \G(0,1,0;y)
          - 3 \G(1,0;y)
         \Big]
       +  \frac{z^2}{4(y+z)^2} \Big[
          - 2\pi^2
          + 3 \H(0;z)
\nonumber \\ &&
          - 12 \H(0;z) \G(0;y)
          - 12 \H(1,0;z)
          - 3 \G(0;y)
          + 12 \G(1,0;y)
         \Big]
       +  \frac{z^2}{4(y+z)} \Big[
          - \pi^2 
          +  \H(0;z)
\nonumber \\ &&
          - 6 \H(0;z) \G(0;y)
          - 6 \H(1,0;z)
          -  \G(0;y)
          + 6 \G(1,0;y)
         \Big]
       +  \frac{z^2}{12(1-y-z)^2}  \Big[
            \frac{5\pi^2}{6}
          + 5 \H(0;z) \G(0;y)
\nonumber \\ &&
          + 5 \H(1,0;z)
          - 5 \G(1,0;y)
         \Big]
       +  \frac{z^2}{4(1-y-z)}  \Big[
          - \pi^2
          +  \H(0;z)
          - 6 \H(0;z) \G(0;y)
          - 6 \H(1,0;z)
          -  \G(0;y)
\nonumber \\ &&
          + 6 \G(1,0;y)
         \Big] 
       +  \frac{z^3}{2(y+z)^4} \Big[
          - \frac{\pi^2}{2} \H(1;z)
          + \frac{\pi^2}{2} \G(1 - z;y)
          + 3 \H(0;z) \G(1 - z,0;y)
          + 3 \H(0,1,0;z)
\nonumber \\ &&
          + 3 \H(1,0;z) \G(1 - z;y)
          - 3 \H(1,0;z) \G(0;y)
          - 3 \H(1,1,0;z)
          - 3 \G(1 - z,1,0;y)
          - 3 \G(0,1,0;y)
         \Big]
\nonumber \\ &&
       +  \frac{z^3}{2(y+z)^3}  \Big[
            \frac{\pi^2}{2}
          + 3 \H(0;z) \G(0;y)
          + 3 \H(1,0;z)
          - 3 \G(1,0;y)
         \Big]
       +  \frac{z^3}{4(y+z)^2} \Big[
            \frac{\pi^2}{2}
          + 3 \H(0;z) \G(0;y)
\nonumber \\ &&
          + 3 \H(1,0;z)
          - 3 \G(1,0;y)
         \Big]
       +  \frac{z^3}{2(y+z)} \Big[
            \frac{\pi^2}{6}
          +  \H(0;z) \G(0;y)
          +  \H(1,0;z)
          -  \G(1,0;y)
         \Big]
\nonumber \\ &&
       +  \frac{z^3}{4(1-y-z)^2} \Big[
          - \frac{\pi^2}{6}
          -  \H(0;z) \G(0;y)
          -  \H(1,0;z)
          +  \G(1,0;y)
         \Big]
       +  \frac{z^3}{2(1-y-z)} \Big[
            \frac{\pi^2}{6}
          +  \H(0;z) \G(0;y)
\nonumber \\ &&
          +  \H(1,0;z)
          -  \G(1,0;y)
         \Big]
       - \frac{1}{12(1-y)} \G(0;y)
       +  \frac{1}{72(1-z)}  \Big[
          - \pi^2
          - 25 \H(0;z)
          + \frac{3}{2} \H(0;z) \G(0;y)
\nonumber \\ &&
          + 9 \H(0,0;z)
          - 6 \H(1,0;z)
         \Big]
       +  \frac{1}{2(y+z)} \Big[
            \frac{\pi^2}{6} \H(1;z)
          - \frac{\pi^2}{6} \G(1 - z;y)
          -  \H(0;z) \G(1 - z,0;y)
          -  \H(0,1,0;z)
\nonumber \\ &&
          - 2 \H(1,0;z)
          -  \H(1,0;z) \G(1 - z;y)
          +  \H(1,0;z) \G(0;y)
          +  \H(1,1,0;z)
          +  \G(1 - z,1,0;y)
          +  \G(0,1,0;y)
\nonumber \\ &&
          - 2 \G(1,0;y)
         \Big]
       +  \frac{1}{6(1-y-z)}  \Big[
          - \frac{\pi^2}{6}
          -  \H(0;z) \G(0;y)
          -  \H(1,0;z)
          -  \G(0;y)
          +  \G(1,0;y)
         \Big]
\nonumber \\ &&
       +  \frac{\pi^2}{72}  \Big[
            \frac{395}{12}
          -  \H(0;z)
          + 2 \G(1 - z;y)
          -  \G(0;y)
          - 2 \G(1;y)
         \Big]
       - \frac{19}{36}\zeta_3
       + \frac{1}{144}  \Big[
            \frac{2977}{18}
          - 10 \H(0;z)
\nonumber \\ &&
          + 24 \H(0;z) \G(1 - z,0;y)
          + 29 \H(0;z) \G(0;y)
          - 36 \H(0;z) \G(0,0;y)
          + 6 \H(0;z) \G(1,0;y)
          - 28 \H(0,0;z)
\nonumber \\ &&
          - 36 \H(0,0;z) \G(0;y)
          - 6 \H(0,1,0;z)
          + 40 \H(1,0;z)
          + 24 \H(1,0;z) \G(1 - z;y)
          - 30 \H(1,0;z) \G(0;y)
\nonumber \\ &&
          - 36 \H(1,0,0;z)
          - 24 \G(1 - z,1,0;y)
          + 56 \G(0;y)
          - 82 \G(0,0;y)
          + 6 \G(0,1,0;y)
          - 40 \G(1,0;y)
\nonumber \\ &&
          + 36 \G(1,0,0;y)
          + 24 \G(1,1,0;y)
         \Big] 
\nonumber \\ &&
 + i\pi \Bigg\{
       - \frac{1}{8(1-z)} \H(0;z)
       + \frac{1}{48} \Big[ 
           \frac{8\pi^2}{3}
          - \frac{46}{3}
          + 13 \H(0;z)
          + 12 \H(0;z) \G(0,y)
          + 12 \H(1,0;z)
          + 31 \G(0;y)
\nonumber \\ && \hspace{9mm}
          - 12 \G(1,0;y)
          \Big]\Bigg\}
\nonumber \\
\lefteqn{A_{\gamma}(y,z) =} \nonumber \\ &&
       +  \frac{1}{48y(1-z)}  \Big[
          - \pi^2
          + 13 \H(0;z)
          - 6 \H(1,0;z)
          - 6 \G(1,0;y)
         \Big]
       +  \frac{1}{48y(1-y-z)}  \Big[
            \pi^2
          - 13 \H(0;z)
\nonumber \\ &&
          + 6 \H(1,0;z)
          + 6 \G(1,0;y)
         \Big]
       +  \frac{z}{16(y+z)^2}  \Big[
          - \frac{11\pi^2}{3} \H(1;z)
          + \frac{11\pi^2}{3} \G(1 - z;y)
          + 22 \H(0;z) \G(1 - z,0;y)
\nonumber \\ &&
          + 22 \H(0,1,0;z)
          + 55 \H(1,0;z)
          + 22 \H(1,0;z) \G(1 - z;y)
          - 22 \H(1,0;z) \G(0;y)
          - 22 \H(1,1,0;z)
\nonumber \\ &&
          - 22 \G(1 - z,1,0;y)
          - 22 \G(0,1,0;y)
          + 55 \G(1,0;y)
         \Big]
       +  \frac{z}{16(y+z)}  \Big[
            \frac{11\pi^2}{3}
          - 11
          - 22 \H(0;z)
\nonumber \\ &&
          + 22 \H(0;z) \G(0;y)
          + 22 \H(1,0;z)
          + 33 \G(0;y)
          - 22 \G(1,0;y)
         \Big]
       +  \frac{z}{8(1-y-z)^2} \Big[
            \frac{\pi^2}{3} \H(1;z)
          + \frac{\pi^2}{6} \G(0;y)
\nonumber \\ &&
          - 2\zeta_3
          +  \H(1,0;z) \G(0;y)
          + 2 \H(1,1,0;z)
          +  \G(0,1,0;y)
         \Big]
       +  \frac{z}{48(1-y-z)}  \Big[
            \frac{10\pi^2}{3}
          + 13 \H(0;z)
\nonumber \\ &&
          + 20 \H(0;z) \G(0;y)
          + 20 \H(1,0;z)
          + 20 \G(0;y)
          - 20 \G(1,0;y)
         \Big]
       +  \frac{z^2}{8(y+z)^3} \Big[
            \frac{22\pi^2}{3} \H(1;z)
\nonumber \\ &&
          - \frac{22\pi^2}{3} \G(1 - z;y)
          - 44 \H(0;z) \G(1 - z,0;y)
          - 44 \H(0,1,0;z)
          - 33 \H(1,0;z)
          - 44 \H(1,0;z) \G(1 - z;y)
\nonumber \\ &&
          + 44 \H(1,0;z) \G(0;y)
          + 44 \H(1,1,0;z)
          + 44 \G(1 - z,1,0;y)
          + 44 \G(0,1,0;y)
          - 33 \G(1,0;y)
         \Big]
\nonumber \\ &&
       +  \frac{z^2}{16(y+z)^2} \Big[
          - \frac{44\pi^2}{3}
          + 33 \H(0;z)
          - 88 \H(0;z) \G(0;y)
          - 88 \H(1,0;z)
          - 33 \G(0;y)
          + 88 \G(1,0;y)
         \Big]
\nonumber \\ &&
       +  \frac{z^2}{16(y+z)}  \Big[
          - \frac{22\pi^2}{3}
          + 11 \H(0;z)
          - 44 \H(0;z) \G(0;y)
          - 44 \H(1,0;z)
          - 11 \G(0;y)
          + 44 \G(1,0;y)
         \Big]
 \nonumber \\ &&
       +  \frac{z^2}{12(1-y-z)^2}  \Big[
            \frac{5\pi^2}{6}
          + 5 \H(0;z) \G(0;y)
          + 5 \H(1,0;z)
          - 5 \G(1,0;y)
         \Big]
       +  \frac{z^2}{16(1-y-z)} \Big[
          - \frac{22\pi^2}{3}
\nonumber \\ &&
          + 11 \H(0;z)
          - 44 \H(0;z) \G(0;y)
          - 44 \H(1,0;z)
          - 11 \G(0;y)
          + 44 \G(1,0;y)
         \Big]
       +  \frac{z^3}{8(y+z)^4}  \Big[
          - \frac{11\pi^2}{2}\H(1;z)
\nonumber \\ &&
          +   \frac{11\pi^2}{2} \G(1 - z;y)
          + 33 \H(0;z) \G(1 - z,0;y)
          + 33 \H(0,1,0;z)
          + 33 \H(1,0;z) \G(1 - z;y)
\nonumber \\ &&
          - 33 \H(1,0;z) \G(0;y)
          - 33 \H(1,1,0;z)
          - 33 \G(1 - z,1,0;y)
          - 33 \G(0,1,0;y)
         \Big]
       +  \frac{z^3}{8(y+z)^3} \Big[
            \frac{11\pi^2}{2}
\nonumber \\ &&
          + 33 \H(0;z) \G(0;y)
          + 33 \H(1,0;z)
          - 33 \G(1,0;y)
         \Big]
       +  \frac{z^3}{16(y+z)^2}  \Big[
            \frac{11\pi^2}{2}
          + 33 \H(0;z) \G(0;y)
          + 33 \H(1,0;z)
\nonumber \\ &&
          - 33 \G(1,0;y)
         \Big]
       +  \frac{z^3}{8(y+z)}  \Big[
            \frac{11\pi^2}{6}
          + 11 \H(0;z) \G(0;y)
          + 11 \H(1,0;z)
          - 11 \G(1,0;y)
         \Big]
\nonumber \\ &&
       +  \frac{z^3}{16(1-y-z)^2}  \Big[
          - \frac{11\pi^2}{6}
          - 11 \H(0;z) \G(0;y)
          - 11 \H(1,0;z)
          + 11 \G(1,0;y)
         \Big]
\nonumber \\ &&
       +  \frac{z^3}{8(1-y-z)} \Big[
            \frac{11\pi^2}{6}
          + 11 \H(0;z) \G(0;y)
          + 11 \H(1,0;z)
          - 11 \G(1,0;y)
         \Big]
       +  \frac{1}{48(1-z)^2} \Big[
            \pi^2 \big( \frac{1}{6} 
          - 3 \H(0;z)
\nonumber \\ &&
          - 3 \H(1;z)
          +  \G(1 - z;y)
          -  \G(0;y) \big)
          - 6\zeta_3 
          + \frac{355}{6} \H(0;z)
          + 6 \H(0;z) \G(1 - z,0;y)
          - 10 \H(0;z) \G(0;y)
\nonumber \\ &&
          - 45 \H(0,0;z)
          - 12 \H(0,0;z) \G(0;y)
          - 18 \H(0,1,0;z)
          +  \H(1,0;z)
          + 6 \H(1,0;z) \G(1 - z;y)
\nonumber \\ &&
          - 6 \H(1,0;z) \G(0;y)
          - 12 \H(1,0,0;z)
          - 18 \H(1,1,0;z)
          - 6 \G(1 - z,1,0;y)
          + 6 \G(0,1,0;y)
         \Big]
\nonumber \\ &&
       +  \frac{1}{48(1-z)}  \Big[
          \pi^2 \big( - \frac{7}{6}
          + 3 \H(0;z)
          + 3 \H(1;z)
          -  \G(1 - z;y)
          +  \G(0;y) \big)
          + 6\zeta_3
          + \frac{277}{6}
          - \frac{493}{6} \H(0;z)
\nonumber \\ &&
          - 6 \H(0;z) \G(1 - z,0;y)
          + 4 \H(0;z) \G(0;y)
          + 45 \H(0,0;z)
          + 12 \H(0,0;z) \G(0;y)
          + 18 \H(0,1,0;z)
\nonumber \\ &&
          - 7 \H(1,0;z)
          - 6 \H(1,0;z) \G(1 - z;y)
          + 6 \H(1,0;z) \G(0;y)
          + 12 \H(1,0,0;z)
          + 18 \H(1,1,0;z)
\nonumber \\ &&
          + 6 \G(1 - z,1,0;y)
          - 10 \G(0;y)
          - 6 \G(0,1,0;y)
          + 6 \G(1,0;y)
         \Big]
       +  \frac{1}{48(1-y-z)} \Big[
          - \pi^2
          + 13 \H(0;z)
\nonumber \\ &&
          - 6 \H(1,0;z)
          - 6 \G(1,0;y)
         \Big]
       +  \frac{1}{48} \Big[
            \pi^2
          - \frac{277}{6}
          + 23 \H(0;z)
          + 6 \H(0;z) \G(0;y)
          + 6 \H(1,0;z)
          + 10 \G(0;y)
\nonumber \\ &&
          - 6 \G(1,0;y)
         \Big] 
 + i\pi \Bigg\{ 
         \frac{11}{16(1-z)^2}\H(0;z)
       + \frac{11}{16(1-z)} \Big[ 1 -  \H(0;z) \Big]  
       - \frac{11}{16} \Bigg\}\nonumber\\
\lefteqn{D_{\gamma}(y;z) =} \nonumber \\&&
       -  \frac{1}{12y(1-z)}\H(0;z)
       +  \frac{1}{12y(1-y-z)} \H(0;z)
       +  \frac{z}{4(y+z)^2} \Big[
            \frac{\pi^2}{3} \H(1;z)
          - \frac{\pi^2}{3} \G(1 - z;y)
\nonumber \\ &&
          - 2 \H(0;z) \G(1 - z,0;y)
          - 2 \H(0,1,0;z)
          - 5 \H(1,0;z)
          - 2 \H(1,0;z) \G(1 - z;y)
          + 2 \H(1,0;z) \G(0;y)
\nonumber \\ &&
          + 2 \H(1,1,0;z)
          + 2 \G(1 - z,1,0;y)
          + 2 \G(0,1,0;y)
          - 5 \G(1,0;y)
         \Big]
       +  \frac{z}{4(y+z)}  \Big[
          - \frac{\pi^2}{3}
          + 1
          + 2 \H(0;z)
\nonumber \\ &&
          - 2 \H(0;z) \G(0;y)
          - 2 \H(1,0;z)
          - 3 \G(0;y)
          + 2 \G(1,0;y)
         \Big]
       +  \frac{z}{12(1-y-z)} \Big[
          - \frac{\pi^2}{3}
          -  \H(0;z)
\nonumber \\ &&
          - 2 \H(0;z) \G(0;y)
          - 2 \H(1,0;z)
          - 2 \G(0;y)
          + 2 \G(1,0;y)
         \Big]
       +  \frac{z^2}{2(y+z)^3} \Big[
          - \frac{2\pi^2}{3} \H(1;z)
          + \frac{2\pi^2}{3} \G(1 - z;y)
\nonumber \\ &&
          + 4 \H(0;z) \G(1 - z,0;y)
          + 4 \H(0,1,0;z)
          + 3 \H(1,0;z)
          + 4 \H(1,0;z) \G(1 - z;y)
          - 4 \H(1,0;z) \G(0;y)
\nonumber \\ &&
          - 4 \H(1,1,0;z)
          - 4 \G(1 - z,1,0;y)
          - 4 \G(0,1,0;y)
          + 3 \G(1,0;y)
         \Big]
       +  \frac{z^2}{4(y+z)^2}  \Big[
            \frac{4\pi^2}{3}
          - 3 \H(0;z)
\nonumber \\ &&
          + 8 \H(0;z) \G(0;y)
          + 8 \H(1,0;z)
          + 3 \G(0;y)
          - 8 \G(1,0;y)
         \Big]
       +  \frac{z^2}{4(y+z)}  \Big[
            \frac{2\pi^2}{3}
          - \H(0;z)
          + 4 \H(0;z) \G(0;y)
\nonumber \\ &&
          + 4 \H(1,0;z)
          +  \G(0;y)
          - 4 \G(1,0;y)
         \Big]
       +  \frac{z^2}{6(1-y-z)^2} \Big[
          - \frac{\pi^2}{6}
          -  \H(0;z) \G(0;y)
          -  \H(1,0;z)
          +  \G(1,0;y)
         \Big]
\nonumber \\ &&
       +  \frac{z^2}{4(1-y-z)} \Big[
            \frac{2\pi^2}{3}
          -  \H(0;z)
          + 4 \H(0;z) \G(0;y)
          + 4 \H(1,0;z)
          +  \G(0;y)
          - 4 \G(1,0;y)
         \Big]
\nonumber \\ &&
       +  \frac{z^3}{2(y+z)^4}  \Big[
           \frac{\pi^2}{2} \H(1;z)
          - \frac{\pi^2}{2}   \G(1 - z;y)
          - 3 \H(0;z) \G(1 - z,0;y)
          - 3 \H(0,1,0;z)
          - 3 \H(1,0;z) \G(1 - z;y)
\nonumber \\ &&
          + 3 \H(1,0;z) \G(0;y)
          + 3 \H(1,1,0;z)
          + 3 \G(1 - z,1,0;y)
          + 3 \G(0,1,0;y)
         \Big]
       +  \frac{z^3}{2(y+z)^3}  \Big[
          - \frac{\pi^2}{2}
\nonumber \\ &&
          - 3 \H(0;z) \G(0;y)
          - 3 \H(1,0;z)
          + 3 \G(1,0;y)
         \Big]
       +  \frac{z^3}{4(y+z)^2} \Big[
          - \frac{\pi^2}{2}
          - 3 \H(0;z) \G(0;y)
          - 3 \H(1,0;z)
\nonumber \\ &&
          + 3 \G(1,0;y)
         \Big]
       +  \frac{z^3}{2(y+z)}  \Big[
          - \frac{\pi^2}{6}
          - \H(0;z) \G(0;y)
          - \H(1,0;z)
          + \G(1,0;y)
         \Big]
       +  \frac{z^3}{4(1-y-z)^2} \Big[
            \frac{\pi^2}{6}
\nonumber \\ &&
          +  \H(0;z) \G(0;y)
          +  \H(1,0;z)
          -  \G(1,0;y)
         \Big]
       +  \frac{z^3}{2(1-y-z)}  \Big[
          - \frac{\pi^2}{6}
          -  \H(0;z) \G(0;y)
          -  \H(1,0;z)
          +  \G(1,0;y)
         \Big]
\nonumber \\ &&
       +  \frac{1}{72(1-z)^2}  \Big[
          - \pi^2
          - 25 \H(0;z)
          + \frac{3}{2} \H(0;z) \G(0;y)
          + 9 \H(0,0;z)
          - 6  \H(1,0;z)
         \Big]
       +  \frac{1}{144(1-z)}  \Big[
            2\pi^2
          - 38
\nonumber \\ &&
          + 65 \H(0;z)
          - 3 \H(0;z) \G(0;y)
          - 18 \H(0,0;z)
          + 12 \H(1,0;z)
          + 3 \G(0;y)
         \Big]
       - \frac{1}{12(1-y-z)}  \H(0;z)
          + \frac{19}{72}
\nonumber \\ &&
          - \frac{5}{48} \H(0;z)
          - \frac{1}{48} \G(0;y)
+ i\pi \Bigg\{ 
       -  \frac{1}{8(1-z)^2}\H(0;z)
       + \frac{1}{8(1-z)} \Big[ -1 +  \H(0;z) \Big]  
       + \frac{1}{8} \Bigg\}
\end{eqnarray}

From the 
$\Omega^{(1),{\rm finite}}$
and 
$\Omega^{(2),{\rm finite}}$, it is possible to recover the 
finite pieces of the helicity-averaged tree times two-loop 
and one-loop squared 
matrix elements by
squaring (\ref{eq:helamp}):
\begin{eqnarray}
{\cal F}inite^{(2\times 0)} (x,y,z) &=& 8\, V\;
{\cal R}\Bigg[ \frac{(1-y)(1-y-z)}{yz}
\alpha^{(2),{\rm finite}}(y,z)
+ \frac{1-y}{y}\beta^{(2),{\rm finite}}(y,z) \nonumber \\
& & \hspace{1.3cm}
-\gamma^{(2),{\rm finite}}(y,z) 
+ (y\leftrightarrow z)
 \Bigg]\; , \nonumber \\
{\cal F}inite^{(1\times 1)} (x,y,z) &=& 4\, V\;
{\cal R}\Bigg[     (1-y-z)\left(\frac{(1-y-z)}{yz} + \frac{1}{2}\right)
          \,
|\alpha^{(1),{\rm finite}}(y,z)|^2 \nonumber \\
&&  \hspace{1cm} + \left( \frac{1-y-z}{2} + \frac{z}{y} \right)\,
|\beta^{(1),{\rm finite}}(y,z)|^2
+ \left( \frac{1-y-z}{2} + \frac{y}{z} \right)\,
|\gamma^{(1),{\rm finite}}(y,z)|^2 \nonumber \\
&& \hspace{1cm} +
          \left(
          - 3+y+z
          + \frac{2-2z}{y}
          \right)\;
\alpha^{(1),{\rm finite}}(y,z)\beta^{*(1),{\rm finite}}(y,z) 
\nonumber \\
&&  \hspace{1cm} - (1-y-z)\; 
 \alpha^{(1),{\rm finite}}(y,z)\gamma^{*(1),{\rm finite}} (y,z)
\nonumber \\
&& \hspace{1cm}
 - (1+y+z)\;  \beta^{(1),{\rm finite}}(y,z)\gamma^{*(1),{\rm finite}}(y,z) 
  + (y\leftrightarrow z) \Bigg]\;.
\label{eq:recover}
\end{eqnarray}

It is important to notice that (\ref{eq:recover}) corresponds, by the 
very nature of the Weyl--van der Waerden helicity formalism, to 
a scheme with external 
momenta and polarization vectors 
 in four dimensions (internal states are 
always taken to be $\dd$-dimensional), 
which is sometimes called the 't Hooft--Veltman 
scheme~\cite{hv}. This scheme is different from the conventional 
dimensional regularization used in~\cite{3jme}, where all external 
momenta and polarization vectors are $\dd$-dimensional. Nevertheless, 
one obtains from (\ref{eq:recover}) the same 
${\cal F}inite^{(2\times 0)} (x,y,z)$ as in Eq.~(4.17) 
and 
${\cal F}inite^{(1\times 1)} (x,y,z)$ as in Eq.~(4.25) 
of~\cite{3jme},
since
all scheme-dependent terms are correctly accounted for by the 
finite contributions arising from expanding the tree level and 
one-loop contributions in the renormalization and infrared factorization 
formulae.  

It should also be noted that only the ${\cal O}(\e^0)$ terms 
of $\Omega^{(1),{\rm finite}}$ contribute to 
${\cal F}inite^{(1\times 1)} (x,y,z)$, terms subleading in 
$\e$ are not required, since no term is multiplied with a divergent factor. 
Comparing the size of these ${\cal O}(\e^0)$ terms (\ref{eq:omega1}) with 
the size of  ${\cal F}inite^{(1\times 1)} (x,y,z)$ in~\cite{3jme}, it 
becomes clear that the squared one-loop amplitude can be evaluated 
much more elegantly by squaring the finite remainders of the 
helicity amplitudes than by computing the 
squared matrix element.

\section{Conclusions and Outlook}
\label{sec:conc}
\setcounter{equation}{0}

In this paper, we have presented analytic formulae for the one- and  two-loop
virtual helicity amplitudes to the  process $\gamma^*\to q\bar q g$.    These
amplitudes have been derived by defining projectors, which isolate the
coefficients of the most general  tensorial structure of the matrix element at
any order in perturbation theory.   Once the general tensor is known, the
helicity amplitudes follow in a straightforward manner -- they are linear
combinations of the tensor coefficients.    We applied the projectors directly
to the Feynman diagrams and used the  conventional approach of relating the
ensuing tensor integrals to a basis set of master integrals.  This latter step
is identical to that employed to evaluate the interference of tree- and
two-loop graphs in Ref.~\cite{3jme}, apart from the fact that the projector is
no longer the tree-level amplitude. As anticipated, the finite remainder from
the interference of tree- and two-loop amplitudes can be reconstructed from the
appropriate helicity amplitudes, with the difference between treating the
external states in $d$ dimensions or four dimensions being isolated in the
infrared-singular terms.   

The results presented here therefore complement the earlier calculation of
the interference of tree- and two-loop graphs in Ref.~\cite{3jme}. Knowledge of
the helicity amplitudes allows additional information on the scattering
process.  In particular, observables that require knowledge of the polarization
tensor of the virtual photon, such as oriented event shapes in unpolarized
$e^+e^-$ scattering or event shapes in polarized $e^+e^-$ scattering,  can be
described at two-loop order.    

Similar results can in principle be obtained for $(2+1)$-jet production in deep
inelastic $ep$ scattering or  $(V+1)$-jet production in hadron--hadron collisions.
However,  the rather different domains of convergence of the 
HPLs and 2dHPLs makes this a non-trivial task, which is discussed in a 
separate paper~\cite{ancont}.  
Nevertheless, the helicity approach will provide
information on the direction of the decay leptons in $(V+1)$-jet production
(with or without polarized protons).   Determination of 
the polarized parton distribution functions in polarized electron--proton 
scattering will also benefit from the
 knowledge of the two-loop helicity amplitudes
in the appropriate kinematic region.

Even though the evaluation of two-loop QCD matrix elements is becoming well
established, the virtual corrections form only part of a full NNLO 
calculation.   They must be combined with the one-loop corrections  to
$\gamma^*\to 4$~partons~\cite{onel4p1,onel4p2,onel4p3,onel4p4},
 where one of the
partons becomes collinear or  soft, as well as  tree-level processes
$\gamma^*\to 5$~partons~\cite{tree5p1,tree5p2,tree5p3} with two soft or
collinear partons in a way that allows all of the infrared singularities to
cancel one another.  This task is far from trivial, even though  the
factorization properties of both the one-loop, one-unresolved-parton
contribution~\cite{onel1,onel2,onel3,onel4,onel5,onel6} and the tree-level, 
two-unresolved-parton contributions~\cite{twot1,twot2,twot3,twot4} have been
studied. Early studies for the case of  photon-plus-one-jet final  states in
electron--positron annihilation in~\cite{ggamma0,ggamma}, which involves both
double radiation and single radiation from one-loop graphs, indicate the
feasibility of developing a numerical NNLO program implementing the 
experimental definition of jet observables and event-shape variables, and
significant progress is anticipated in the near future. 
\vspace{4mm}

{\bf Note added:} After this paper was first released, part of its results 
were confirmed in an independent calculation using the methods described 
in~\cite{muw,w2}. In hep-ph/0207043, Moch, Uwer and Weinzierl  
obtain results for the full one-loop amplitude (\ref{eq:oneloopamp})
and for the contributions to the two-loop amplitude (\ref{eq:twoloopamp})
which are proportional to $N_F$ (i.e.\ the terms $D_\Omega$ and $E_\Omega$),
all in agreement with the results presented here.

\section*{Acknowledgements} 
EWNG thanks Adrian Signer for useful discussions.
This work was supported in part by the EU Fourth Framework Programme 
``Training and Mobility of Researchers'', 
network ``Quantum Chromodynamics and the Deep Structure of 
Elementary Particles'', contract FMRX-CT98-0194 (DG 12-MIHT).

\begin{appendix}

\renewcommand{\theequation}{\mbox{\Alph{section}.\arabic{equation}}}

\section{Weyl--van der Waerden spinor calculus}
\setcounter{equation}{0}
\label{sec:appa}

The basic quantity is the two-spinor $\psi_{A}$ or $\psi^A$
and its complex conjugate
$\psi_{\dot{A}}$ or $\psi^{\dot{A}}$. Raising and lowering of indices is
done with the antisymmetric tensor $\varepsilon$,
\begin{eqnarray}
  \varepsilon_{AB} = \varepsilon^{AB} = \varepsilon_{\dot{A}\dot{B}} =
\varepsilon^{\dot{A}\dot{B}} =
  \left( \begin{array}{cc} 0 & 1 \\ -1 & 0 \\ \end{array}\right) .
\end{eqnarray}
We define a antisymmetric spinorial ``inner product'':
\begin{equation}
 \langle\psi_{1}\psi_{2}\rangle = \psi_{1A}\varepsilon^{BA}\psi_{2B}
  = \psi_{1A}\psi^{A}_{2} = -\psi_1^A\psi_{2A}= -\langle\psi_{2}\psi_{1}\rangle,
\end{equation}
and
\begin{equation}
\langle\psi_{1}\psi_{2}\rangle^* = \psi_{1\dot{A}}\psi^{\dot{A}}_{2} .
\end{equation}
Any momentum vector $k_{\mu}$ gets a bispinor representation by
contraction with $\sigma^{\mu}$:
\begin{equation}
k_{\dot A B}=\sigma^\mu_{\dot A B}k_\mu=
\left(\begin{array}{cc}
k_0+k_3 & k_1+ik_2 \\ k_1-ik_2 & k_0-k_3
\end{array}\right),
\end{equation}
where $\sigma^{0}$ is the unit matrix and $\sigma_{i}$ are the Pauli
matrices.
Since
\begin{equation}
  \sigma^{\mu}_{\dot{A}B}\sigma^{\nu\dot{A}B} = 2g^{\mu\nu},
\end{equation}
we have
\begin{equation}
  k_{\dot{A}B}p^{\dot{A}B} = 2k\cdot p .
\end{equation}
For light-like vectors one can show that
\begin{equation}
  k_{\dot{A}B} = k_{\dot{A}}k_{B},
\end{equation}
where
\begin{equation}
k_A =\left(\begin{array}{c}
(k_1-ik_2)/\sqrt{k_0-k_3} \\ \sqrt{k_0-k_3}
\end{array}\right),
\end{equation}
so that for light-like vectors we have
\begin{equation}
2k\cdot p =\langle kp\rangle\langle kp\rangle^*=\left|\langle kp \rangle\right|^2 .
\end{equation}
The following relation is often useful:
\begin{equation}
\sigma^\mu_{\dot A B}\sigma_\mu^{\dot C D}=2{\delta_{\dot A}}^{\dot C}
{\delta_B}^D\ .
\end{equation}

For massless spin-$\frac{1}{2}$ particles the four-spinors can be
expressed in two-spinors as follows:
\begin{eqnarray}
u_+(p)&=&v_-(p)=
\left(\begin{array}{c}p_B\\0\end{array}\right) ,\nonumber \\
u_-(p)&=&v_+(p)=
\left(\begin{array}{c}0\\p^{\dot B}\end{array}\right) ,\nonumber \\
\bar{u}_+(q)&=&\bar{v}_-(q)=
\left(\begin{array}{cc}0,&-iq_{\dot A}\end{array}\right) ,\nonumber \\
\bar{u}_-(q)&=&\bar{v}_+(q)=
\left(\begin{array}{cc}iq^A,&0\end{array}\right) .
\end{eqnarray}
The Dirac $\gamma$ matrices now become
\begin{equation}
\gamma^\mu=\left(\begin{array}{cc}0&-i\sigma^\mu_{\dot B A} \\
i\sigma^{\mu\dot A B}&0\end{array}\right) ,
\end{equation}
so that, for example:
\begin{equation}
\bar{u}_{+}(q)\gamma^{\mu}v_{-}(p) = q_{\dot{A}}\sigma^{\mu\dot{A}B}p_{B}~.
\end{equation}

The general electroweak vertex for vector boson $V$ coupling to two 
fermions is denoted by 
$ie\delta_{ij}\Gamma_\mu^{Vf_1f_2}$, where $i$ and $j$ are the colour 
labels associated with the fermions $f_1$ and $f_2$ respectively.
The vertex contains left- and right-handed couplings,
\begin{equation}
\Gamma_\mu^{V,f_1f_2} = 
L^V_{f_1f_2}\gamma_\mu\left(\frac{1-\gamma_5}{2}\right)
+R^V_{f_1f_2}\gamma_\mu\left(\frac{1+\gamma_5}{2}\right),
\end{equation}
where for a photon,
\begin{equation}
L^\gamma_{f_1f_2}
=R^\gamma_{f_1f_2}
=-e_{f_1} \delta_{f_1f_2},
\end{equation}
and for a $Z$-boson,
\begin{equation}
L^Z_{f_1f_2}=\frac{I_3^{f_1}-\sin^2\theta_We_{f_1}}
{\sin\theta_W\cos\theta_W}\delta_{f_1f_2},~~~~~~
R^Z_{f_1f_2}
=\frac{-\sin\theta_We_{f_1}}
{\cos\theta_W}\delta_{f_1f_2}.
\end{equation}
Here, $e_f$ represents the fractional electric charge, $I_3^f$ the weak 
isospin and $\theta_W$ the weak mixing angle.  
In the Weyl--van der Waerden notation, the vertex $\Gamma_\mu^{V,f_1f_2}$ 
becomes,
\begin{equation}
\Gamma_\mu^{V,f_1f_2}=
\left(\begin{array}{cc}0&-iL^V_{f_1f_2}\sigma_{\mu\dot B A} \\
iR^V_{f_1f_2}\sigma_\mu^{\dot A B}&0\end{array}\right) .
\end{equation}
 
For the polarization vectors of outgoing gluons and photons we use the
spinorial quantities
\begin{eqnarray}
  e^{+}_{\dot{A}B}(k) & = & \sqrt{2} \frac{k_{\dot{A}}b_{B}}{\langle bk\rangle}~, \\
  e^{-}_{\dot{A}B}(k) & = & \sqrt{2} \frac{b_{\dot{A}}k_{B}}{\langle bk\rangle^{*}}~.
\end{eqnarray}
The gauge spinor $b$ is arbitrary and can be chosen differently in each
gauge-invariant expression. A suitable choice can often 
simplify the calculation.

\end{appendix}

\end{document}